\documentclass[linenumbers]{aastex631}

\usepackage{natbib}
\usepackage{booktabs}

\usepackage{graphicx,float,amsmath,bm,epsfig,epstopdf}

\shortauthors{Pal et al.}

\begin{document}
\nolinenumbers
\title{Automatic detection of large-scale flux ropes and their geoeffectiveness with a machine learning approach}

\author{Sanchita Pal}
\affiliation{NASA Postdoctoral Program Fellow, NASA Goddard Space Flight Center, Greenbelt, MD 20771, USA}
% \affiliation{Heliophysics Science Division, NASA Goddard Space Flight Center, Greenbelt, MD 20771, USA}

\author{Luiz F. G. dos Santos}
\affiliation{SETI Institute, Mountain View, CA 94043, USA}
\author{Andreas J. Weiss}
\affiliation{NASA Postdoctoral Program Fellow, NASA Goddard Space Flight Center, Greenbelt, MD 20771, USA}
\author{Thomas Narock}
\affiliation{Center for Natural, Computer, and Data Sciences, Goucher College, Baltimore, MD, United States}
\author{Ayris Narock}
\affiliation{ADNET Systems Inc., 7515 Mission Drive, Lanham, MD 20706, USA}
\affiliation{Heliophysics Science Division, NASA Goddard Space Flight Center, Greenbelt, MD 20771, USA}
\author{Teresa Nieves-Chinchilla}
\affiliation{Heliophysics Science Division, NASA Goddard Space Flight Center, Greenbelt, MD 20771, USA}
\author{Lan K. Jian}
\affiliation{Heliophysics Science Division, NASA Goddard Space Flight Center, Greenbelt, MD 20771, USA}
\author{Simon W. Good}
\affiliation{Department of Physics, University of Helsinki, P.O. Box 64, FI-00014 Helsinki, Finland}
%%%%%%%%%%%%%%%%%%%%%%%%%

\begin{abstract}
Detecting large-scale flux ropes (FRs) embedded in interplanetary coronal mass ejections (ICMEs) and assessing their geoeffectiveness are essential since they can drive severe space weather. At 1 au, these FRs have an average duration of 1 day. Their most common magnetic features are large, smoothly rotating magnetic fields. Their manual detection has become a relatively common practice over decades, although visual detection can be time-consuming and subject to observer bias. Our study proposes a pipeline that utilizes two supervised binary-classification machine learning (ML) models trained with solar wind magnetic properties to automatically detect large-scale FRs and additionally determine their geoeffectiveness. The first model is used to generate a list of auto-detected FRs. Using the properties of southward magnetic field the second model determines the geoeffectiveness of FRs. Our method identifies 88.6\% and 80\% large-scale ICMEs (duration $\ge 1$ day) observed at 1 au by Wind and Sun Earth Connection Coronal and Heliospheric Investigation (STEREO) mission, respectively. While testing with a continuous solar wind data obtained from Wind, our pipeline detected 56 of the 64 large-scale ICMEs during 2008 - 2014 period (recall= 0.875) but many false positives (precision= 0.56) as we do not take into account any additional solar wind properties than the magnetic properties. We found an accuracy of 0.88 when estimating the geoeffectiveness of the auto-detected FRs using our method. Thus, in space weather now-casting and forecasting at L1 or any planetary missions, our pipeline can be utilized to offer a first-order detection of large-scale FRs and geoeffectiveness. 

\end{abstract}


\section{Introduction} \label{sec:1}
Magnetic flux ropes (FRs) are fundamental solar and heliospheric structures that are formed by twisted bundles of magnetic field lines. They can originate at the Sun or in the heliosphere. They are commonly associated with coronal mass ejections \citep[CMEs;][] {1997GMS....99.....C, Webb_Howard2012}, streamer blow-outs \citep[SBOs;][]{Vourlidas2018,Nitta2021}, small transients \citep[STs;][]{2009SoPh..256..327K,2014JGRA..119..689Y,2016JGRA..121.5005Y} in solar wind, local small-scale magnetic flux rope structures near the heliospheric current sheet (HCS) \citep{1995JGR...10019903M,2000GeoRL..27...57M} and magnetospheric flux transfer events \citep[FTEs;][]{2020EPSC...14...62S}. 
\par 
The large-scale FRs of interplanetary counterparts of CMEs (ICMEs) may have a significant contribution of flux from coronal reconnection \citep{2007ApJ...659..758Q, 2017SoPh..292...65G, 2017ApJ...851..123P, 2018ApJ...865....4P, 2022AdSpR..70.1601P} and thus may transport large quantities of solar mass, magnetic flux and helicity away from the Sun. Frequently these FRs are called magnetic clouds \citep[MCs;][]{1981JGR....86.6673B} which have an enhanced magnetic field strength, a smooth and coherent rotation in magnetic field lines \citep[e.g.][]{1981JGR....86.6673B, 1982JGR....87..613K}, declining velocity \citep[e.g.][]{1982JGR....87..613K, 2003SoPh..216..285R}, low proton temperature and low proton $\beta$ (the ratio between proton thermal pressure and magnetic pressure) characteristics \citep[e.g.][]{1995JGR...10023397R}. At 1 au, ICME FRs have an average radial dimension of $\sim$0.25 au \citep{1982JGR....87..613K}. A duration of approximately 1 day, corresponds to a characteristics dimension $\sim$0.25 au \citep{1982JGR....87..613K, 2007JGRA..112.9103K, 2009AnGeo..27.3479L}. However, studies like \citet{2004GeoRL..3118804R} show that there may be a lack of MCs in ICMEs observed at different heliocentric distances and different phases of the solar cycle. It may be possible that all ICMEs contain a well-defined flux rope close to their origin, but as ICMEs evolve on their way out from the Sun, their flux rope signatures may weaken \citep{1997GMS....99.....C}. Alternatively, ICMEs may always contain identifiable FRs but, when they are traversed far from their central FR axis, their FRs cannot be detected \citep[e.g.][]{2006SoPh..239..393J}. \citet[]{2006SoPh..239..393J} and \citet[]{2018SoPh..293...25N} adopted a less restrictive term called magnetic obstacles (MOs) to address the magnetic structure inside ICMEs. The MOs are described as periods inside ICMEs having low proton $\beta$ where in some cases, the magnetic field lines display smooth and monotonic rotations.  \par
Apart from ICME FRs, small transients (STs) that may last from a few tens of minutes to a few hours are one of the important sources of FRs in the heliosphere. They are components of slow solar wind and may originate from the cusps of coronal streamers \citep{1997ApJ...484..472S, 2008ApJ...675..853S, 2000JGR...10525133W}. \citet{2014JGRA..119..689Y, 2016JGRA..121.5005Y} have explained the ST characteristics in details. They differ from ICME FRs in a number of ways, including their duration (0.5–12 hours), the absence of a speed differential with respect to the slow background solar wind, and the lack of a discernible amplification of the heavy ion charge state within them. Some of the STs may have the geometry of small-scale flux ropes (SFRs) that are local structures mostly observed near HCS and are generated due to magnetic reconnection and a turbulence cascade \citep{2008JGRA..113.9105C, 2010JGRA..115.8102C}. Although SFRs show the typical magnetic configuration of FRs (i.e., smooth rotation) they may have proton $\beta$ less than or greater than 1.  \par
Decades of studies have been conducted to explore how solar magnetic fields affect the geospace environment through the solar wind and solar storms \citep{2023JASTP.24806081N}. The major solar storms including ICMEs may shape the near-Earth heliospheric environment in a way that may trigger the most extreme forms of space weather at Earth. Even a moderate type geomagnetic storm is capable to adversely affect assets in geospace environment \citep{2024SpWea..2203716B}. It is challenging to predict their impact on the Earth as they may continuously evolve in the heliosphere while propagating \citep{2017SSRv..212.1159M, 2020SoPh..295...61L, 2022A&A...665A.110P,2023FrASS..1095805P}. Their geo-effectiveness and capability to trigger magnetic storms have societal impacts, which cannot be disregarded in modern society. With high proton number density, increased proton speed and magnetic field component $B_z$ anti-parallel to the geomagnetic field, ICMEs impact the magnetosphere and also allow solar energetic particles to drift in the magnetosphere. A simplified explanation of ICME - magnetosphere interaction \citep{1961PhRvL...6...47D} states that if an ICME $B_z$ is directed opposite to the Earth’s field, magnetic reconnection takes place at the day-side magnetosphere, and reconnected magnetic flux accumulates in the night side magnetotail region. Thus, the magnetic reconnection at the magnetotail leads to solar plasma injection toward the Earth on the night side. This results in the formation of aurora and a ring current around the Earth, which causes a reduction in the geomagnetic field measured by the disturbance storm time (Dst) index. \par 
The ICME speed $V$ and its southward magnetic field component $-B_z$ are of the most importance in creating geomagnetic storms \citep{1966JGR....71..155F, 2000GeoRL..27.3591C, 2002JGRA..107.1314W}. Certainly, a sufficiently long duration ($\delta t$) of $-B_z$ is also necessary \citep{2003GeoRL..30.2039W}. The product of the average value of $\overline{-VB_z}$ and $\delta t$ defines the magnetic flux transferred from the interplanetary medium into the magnetosphere. \citet{2003GeoRL..30.2039W} performed a statistical study of the relationship between $\overline{-VB_z}$, $\delta t$ and $Dst_{min}$ that revealed the importance of $\overline{-VB_z}$ and its duration $\delta t$ for generating geomagnetic storms. The study found an empirical formula $Dst_{min}=-19.01-8.43(\overline{-VB_z})^{1.09}(\delta t)^{0.3}$ that indicates a more prominent contribution of large $\overline{-VB_z}$ over $\delta t$ to create a strong geomagnetic storm. They found a high anti-correlation with correlation coefficient (CC) of $-0.95$ between ($\overline{-VB_z})^{1.09}(\delta t)^{0.3}$ and $Dst_{min}$, whereas, a comparatively weak linear correlation with CC of $-0.7$ was found between ($\overline{-VB_z})(\delta t$) and $Dst_{min}$. This suggests a non-linear relation between $Dst_{min}$ and $\delta t$, whereas $Dst_{min}$ and $\overline{-VB_z}$ have an almost linear relationship.  \par

Visual inspection of the solar wind magnetic field and plasma environment via spacecraft data to identify ICMEs is widely practiced though it is quite time-consuming, ambiguous, subjective, and biased by the observer's interpretations \citep{2003AGUFMSH21B0133S}. Some studies have begun to introduce the use of supervised machine learning (ML) algorithms that have the potential to learn the properties of labeled events by themselves. Their use in space physics is progressing since they can tackle a large and ever-growing database accumulated over decades. In solar physics, ML-based techniques have recently bloomed as a very attractive way of using our computing resources to extract conclusions from solar data \citep{2023LRSP...20....4A}. \par
% Several studies have taken advantage of ML techniques in the classification of solar wind data. \citet{2017JGRA..12210910C} applied the Gaussian Process that uses a Bayesian approach to categorize solar wind data in multiple classes. This method had a median accuracy larger than 90\% for all categories. 
\citet{2019ApJ...874..145N} proposed a pipeline that used a Convolutional Neural Network (CNN) trained with 33 solar wind magnetic and plasma parameters obtained and derived from in situ spacecraft to automatically detect the start and end times of ICMEs in in-situ data. Their method found 84\% of the listed ICMEs collected from different existing catalogs where ICMEs are manually detected using Wind spacecraft data during 2010--2015. Out of 158 ICMEs found during the period using their pipeline, 25 ($\sim$16\%) events were false positives. \citet{2020SoPh..295..131D} implemented a tool based on handwriting recognition models that uses a deep CNN to classify subsets of the ICMEs introduced in \citet{2019SoPh..294...89N}. The weights of their model were trained with synthetic data obtained from a well-established physical flux rope model. They simulated the noise of synthetic training data using data points belonging to a normal distribution with standard deviations including 0.05 and 0.1. Their methodology classified 76\% of 321 real cases (identified manually by \citet{2019SoPh..294...89N} using Wind spacecraft data) correctly during 1995--2015. \citet{2022FrASS...938442N} attempted to predict the orientation of heliospheric magnetic flux ropes once they are identified in in-situ solar wind data using a model based on CNN architecture. \citet{2022SpWea..2003149R} converted the ICME detection problem to a time series segmentation problem and proposed a pipeline for automatic detection of ICMEs using a variation of an advanced U-Net \citep{10.1007/978-3-319-24574-4_28} method that has recently proven successful in medical image segmentation. Using their method, 73\% of the cataloged events observed by the Wind spacecraft during 1997-2015 were detected correctly. They found a total number of 254 ($\sim$40\%) events as false positives during that period. A comparable method was proposed by \citet{2022ApJS..259....8C} that uses the residual U-net architecture and automatically detected 77\% of ICMEs during the period 1997 October 1 to 2016 January 1 collected by the Wind spacecraft. Based on Grad-Shafranov reconstruction technique, several studies have developed automated SFR detection algorithms \citep{2018ApJS..239...12H}, and applied them to the spacecraft data at different heliocentric distances \citep{2019ApJ...881...58C, 2020ApJ...894...25C, 2022ApJ...924...43C}. Using ML-based classification algorithm \citet{2024ApJ...961...81F} compared SFRs, MCs, and ambient solar wind plasma properties. \par

A recent study by \citet{nair2023magnet} summarises the results obtained from an open data-science challenge called ``MagNet: Model the Geomagnetic Field" to Predict Dst from solar wind Data. The challenge attracted 1,197 model submissions that used various ML approaches and predicted Dst in a real-time modeling environment. \citet{2023SpWea..2103286H} developed a 6-hour ahead Dst prediction model from solar wind observations using Gated Recurrent Unit networks. They used several solar wind parameters like proton density, velocity, magnetic field intensity, magnetic field vector in the z-direction, clock, and dipole tilt angle of the magnetic field, and Dst index and boosted the performance of the model by developing a multi-fidelity method. \citet{2020ApJS..248...14X} used an ensemble-learning algorithm that combines the artificial neural network, support vector regression, and long short-term memory network to predict the Dst index 1–6 hr in advance. They used solar wind's total magnetic field, $B_z$, total electric field, solar wind speed, plasma temperature, and proton pressure as inputs. \citet{2017SpWea..15.1004C} provided a method to generate a probabilistic prediction of Dst using a Gaussian method approach. \citet{2018SpWea..16.1882G} constructed the Dst index prediction model by combining the long short-term memory (LSTM) network with the Gaussian process approach. To forecast the Dst index during 80 intense geomagnetic storms ($Dst_{min}\leq100$ nT) during 1995 - 2014, \citet{2016P&SS..120...48L} applied a support vector machine (SVM) and combined that with a distance correlation technique. They utilized 13 observed and derived plasma and magnetic parameters of the solar wind as inputs. \citet{2022ApJ...934..176P} developed binary classification models to predict the geoeffectiveness of CMEs, i.e. whether a CME is associated with a geomagnetic storm having $Dst_{min}\leq30$ nT or not. They used supervised ML models like logistic regression, k-nearest neighbors, SVM, feed-forward artificial neural networks, and ensemble models trained on white-light coronagraph data sets of CMEs close to the Sun. They achieved an adequate hit rate with these models. \citet{2012JKAS...45...31C} found a better performance of SVM than the performance of simple classification based on constant classifiers, when they used SVM for the prediction of geoeffective CMEs. The SVM algorithm is proven to be efficient in a wide range of solar and space physics research, such as flare classification \citep[e.g.][]{2007SoPh..241..195Q, 2022ApJ...935...45S}, ICME arrival time prediction \citep[e.g.][]{2018ApJ...855..109L}, Dst index prediction \citep[e.g.][]{2016P&SS..120...48L}, and sunspot occurrence prediction \citep[e.g.][]{2022PASP..134l4201R}. \par

Previous studies reveal an impressive performance of ML-based methods in the field of automatic ICME detection and their geoeffectiveness prediction. Integrating automatic detection of both space weather driving events and their geoeffectiveness in a single pipeline is an efficient way for quickly assessing the real-time space weather and issuing alert. In this study, we present a pipeline that automatically detects large-scale ICME associated FRs from continuous spacecraft measurements at 1 au and additionally indicates their geoeffectiveness -- whether they are minor to intensely geoeffective ($Dst_{min}\leq-50$ nT) or not. For the FR detection process, the pipeline uses CNN architecture trained with more realistic synthetic FR and non-flux rope (NFR) magnetic fields prepared using a physics-based FR model combined with power spectral densities of magnetic field fluctuations in FRs and NFRs derived from spacecraft data. For indicating the geoeffectiveness of the automatically detected FRs, our pipeline uses another supervised binary classifier trained with geoeffective magnetic parameters of real events. 
\par 
In Section \ref{sec:2}, we present our proposed pipeline and describe our methodology of pre-processing the dataset, training the models, and post-processing techniques. Section \ref{sec:3} discusses the results and Section \ref{sec:4} summarises and concludes the article.

\section{ Description of pipeline} \label{sec:2}
% We propose a pipeline in order to classify continuous solar wind magnetic field data as flux rope (FR) and non flux rope (NFR) segments, and later their geoeffectiveness. First, w
We combine a supervised ML model `M1' with our current understanding of ICME FR and non flux rope (NFR) solar wind segments and integrate this to another supervised ML model `M2' to first detect the ICME FR intervals from solar wind data and then their geoeffectiveness. This work utilizes only the magnetic properties rather than other solar wind properties like composition and plasma. This is because in the future, we plan to make this pipeline usable in classification of real-time solar wind being observed by various planetary missions. Unlike magnetic field data, plasma and composition data are generally not available in real-time with reliable quality from planetary missions. 
% from planetary missions than plasma data.} 
 
\par
\begin{figure}[!tbp]
  \centering
    \includegraphics[width=1\textwidth]{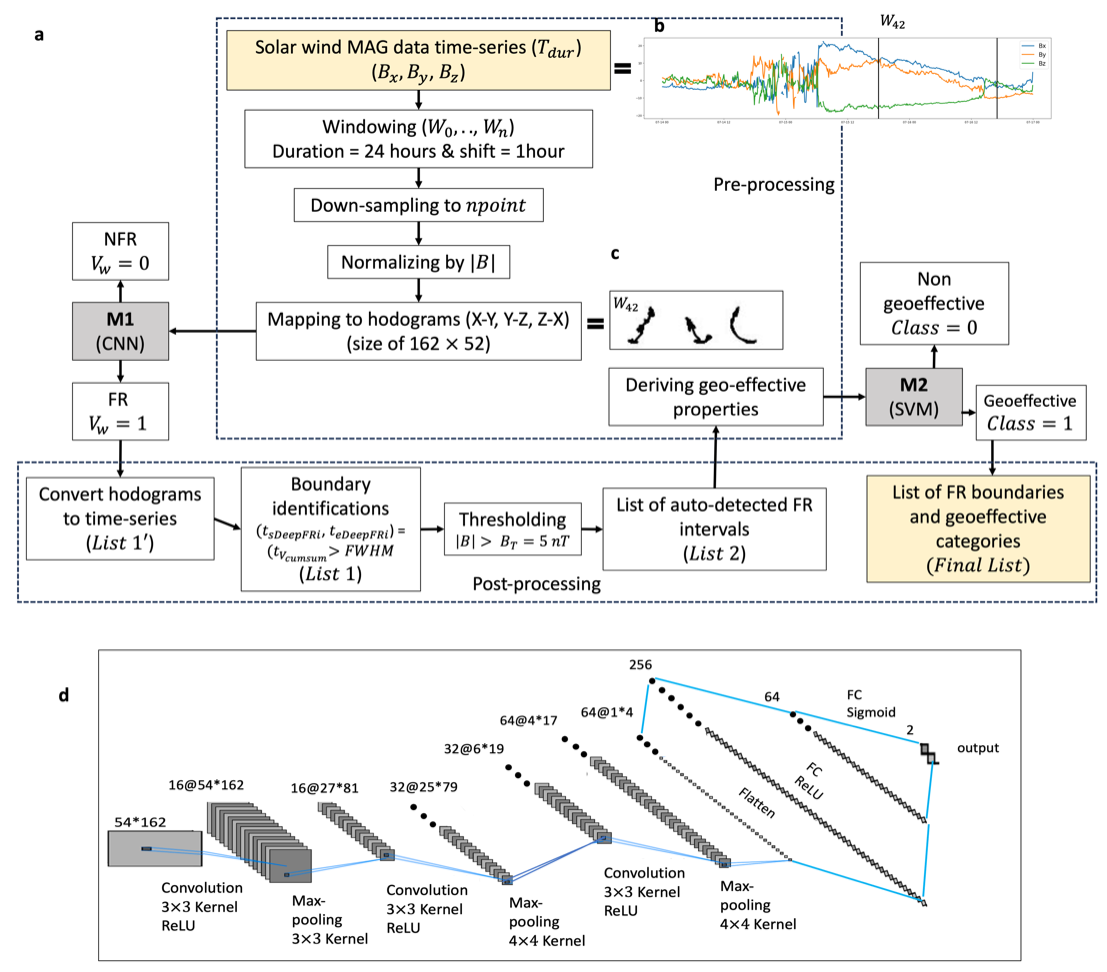}
  \caption{Upper panel- (a) A flowchart of our pipeline that shows the processes followed to convert raw data from the Wind spacecraft to a list of auto-detected FRs and their geoeffectiveness categories. The input and the final output of the flowchart are shown in yellow boxes. The models used are shown in grey boxes. The dashed boxes enclose the pre-processing and post-processing steps. As an example, (b) a 3-day interval Wind magnetic field vector data is shown on the right side where the intervals between black and green vertical lines indicate two consecutive windows ($W_{42}$ and $W_{43}$) of 24 hours duration, respectively. The corresponding merged hodogram is shown at (c). Lower panel- schematic of M1 model architecture. The gray squares represent output of each layers. (d) Schematic of the CNN architecture used for M1 model. The squares represent the output of each layer. The architecture consists of three sets of convolutional and max-pooling layers using ReLU activation, followed by two fully connected layers with ReLU and sigmoid as the activation functions, respectively.  }
  \label{fig1}
\end{figure}

 
In the pipeline, the first model `M1' is an image classification model that uses CNN architecture. It is trained on synthetic large-scale FRs and NFRs prepared by combining a well-established physical flux rope model and a sophisticated additive background solar wind profile model based on the power spectrum density (PSD) of solar wind magnetic field corresponding to real ICME FRs and NFRs. We refer to the model as \textit{DeepFRi} -- \textbf{Deep} convolutional neural network \textbf{FR} \textbf{i}dentification)”. The second model that detects the geoeffectiveness of FRs is named as \textit{DicFR} -- Disturbance storm time (\textbf{D}st) \textbf{i}ndex \textbf{c}lassification of identified \textbf{FR}s. Note that the DicFR model aims to classify auto-identified FRs based on their geoeffectiveness, i.e. whether they are moderate to intensely geoeffective or not. It doesn't intend to predict the Dst index value \citep[e.g.][]{2022SpWea..2003064H}. In the upper panel of Figure \ref{fig1}, using a flowchart we show the operations that we perform in this work to auto-identify FRs and their geoeffectiveness from solar wind data. As a final result, the pipeline generates a list that contains auto-identified FR start and end times, and their geoeffectiveness categories. Below we describe the pre-processing of data, the models, and the post-processing steps including boundary determination of auto-identified FRs and listing them with their predicted geoeffectiveness categories.   \par

% In this Section, we discuss the different steps of the pipeline that enables automatic identification of FRs and their geoeffectiveness, the model architectures, and preparation of training and evaluation data. 
% \subsection{Pipeline}
% In Figure \ref{fig1}, we present a flowchart involving the different steps of the pipeline that shows the pre-processing of data, training DeepFRi and DicFR models and post-processing the model outputs to create catalogs containing auto-detected FRs and predicted geoeffectivity of the detected FRs. The dashed blocks separate different major steps of our pipeline.

\subsection{Data Pre-processing}\label{pre-processing}
As our first model DeepFRi is intended to classify solar wind into FRs and NFRs based on their magnetic characteristics, we use only processed and calibrated Magnetic Field Investigation \citep[MFI][]{1995SSRv...71..207L} instrument data at 1 min resolution during 1995 - 2021 from the Wind spacecraft. The Wind/MFI instrument-measured solar wind magnetic field data in Geocentric Solar Ecliptic (GSE) coordinate system ($x_{gse}, y_{gse}, z_{gse}$). We first window the MFI data in 24-hour intervals having $24\times 60$ data points and check for the data gaps. If there are data gaps within the window and the gaps cover no more than 10\% of the window size the window is still considered as an acceptable input. 
The size of the window is chosen to be 24 hours because we focus here only to detect ICME-associated large-scale FRs and in general, at 1 au, the average ICME FRs are of $\sim$24-hour duration \citep{1982JGR....87..613K, 1998AnGeo..16....1B}. To verify this, we consider 695 manually identified ICME FRs obtained from \href{https://wind.nasa.gov/ICME_catalog/ICME_catalog_viewer.php}{Wind ICME catalog} \citep{2018SoPh..293...25N} and \href{https://stereo-ssc.nascom.nasa.gov/data/ins_data/impact/level3/}{STEREO ICME catalog} \citep{2018ApJ...855..114J} which are observed by both Wind and the Solar TErrestrial RElations Observatory \citep[STEREO;][]{2008SSRv..136....5K} spacecraft, respectively at 1 au and find their mean duration as 24.3 hours. Next, we sample each window into $npoint$, where each point corresponds to $\sim5$-min average of solar wind magnetic field data.
% where $npoint$ is a fixed integer that is less than the number of points found in the present window and decided during the training process of M1. 
After down-sampling the window, we normalize each magnetic field component with the solar wind magnetic field intensity in that window. Next, the normalized magnetic field components of the down-sampled window are plotted in three different planes $x_{gse} - y_{gse}$, $y_{gse} - z_{gse}$ and $z_{gse} - x_{gse}$, where each of them is in the -1 to 1 range and of pixel size (Width) 54 $\times$ (Height) 54. These are called magnetic field hodograms. After generating three different hodogram images, we merge them into a single image of pixel size (Width) 162 $\times$ (Height) 54. Thus, each window corresponds to a merged hodogram acceptable for our M1 model input. \par
If we consider a time interval ($T_{dur}$) starting at $t_s$ hours and ending at $t_e$ hours, we divide the interval in $n$ number of windows -- $W_0, W_1, ...W_{n-1}$, where $n$ is an integer with a unit of hours. Thus, $T_{dur}$ (in hours) can be expressed as,

\begin{equation}
\begin{split}
    T_{dur} = \sum_{n=0}^{(t_e-t_s)+24h} W_n,\\  
    W_n=\int_{t_s-24h+n}^{t_s+n} dt.
\end{split}
\end{equation}
Here,
% $round$(`X') provides the nearest integer of `X' and 
dt is in hour. \par

As an example, in Figure \ref{fig1}, we show a magnetic vector time series ($T_{dur}$). Inside $T_{dur}$, two windows -- $W_n$ and $W_{n+1}$ (n = 42) of 24 hours duration are the intervals between the pairs of black and green vertical lines. A merged hodogram corresponding to $W_{42}$ is also shown which can be directly used as input to M1. 
\par

For the second model M2 that is intended to classify the geoeffectiveness, we extract the geoeffective magnetic properties from Wind/MFI time series of the auto-detected FR intervals in Geocentric Solar Magnetospheric system (GSM). The model is trained using FR geoeffective properties and the minimum value of $Dst$ index during the FR intervals obtained from \href{https://omniweb.gsfc.nasa.gov/form/omni_min.html}{OMNIWeb database}. The geoeffective magnetic characteristics that we consider for this case are the minimum value of the magnetic field component in the $-z_\mathrm{gsm}$ direction ($-B_{z_\mathrm{gsm},min}$) and the duration of $-B_{z\mathrm{gsm}}$ with respect to the whole $B_{z\mathrm{gsm}}$ duration in a window. The second feature is named $R_{Bz}$. To obtain this, we consider only the hodogram in the $y_\mathrm{gsm} - z_\mathrm{gsm}$ plane and find the ratio of the number of points in a plane formed by $y_\mathrm{gsm} - (-z_\mathrm{gsm})$ and the total number of points in the whole $y_\mathrm{gsm} - z_\mathrm{gsm}$ plane. Thus, once we obtain an automated catalog of FRs from model M1, we process each identified FR to obtain $-B_{z_\mathrm{gsm},min}$ and $R_{Bz}$ and provide them as input to the M2 model that is trained to classify FRs as geoeffective or non-geoeffective.\par

\subsection{Models}
In this Section, we describe model architectures, the method of training the models and their evaluations.
% and preparing the list of auto-detected events.
\subsubsection{Model architectures}
The model M1 uses a binary classification CNN architecture that is based on handwriting recognition models. A CNN is a deep learning architecture consisting of interconnected nodes or neurons having learnable filters constructing layers, whose objectives are to detect features and patterns present in a data set. Thus, a certain number of layers process information in a way similar to the human brain \citep{2015Natur.521..436L, Goodfellow-et-al-2016}. In each layer, filters are convolved along the input data, and using a nonlinear activation function the layer indicates the presence (or absence) of the requested pattern. A complete description of the principles of CNN can be found in \citet{geron2017handson}. \par
The architecture of M1 shown in the lower panel of Figure \ref{fig1} is quite similar to that of the model used by \citet{2020SoPh..295..131D} which implemented a binary class handwritten digit-recognition model \citep{10.5555/2283516.2283603}. The input to our model are merged hodograms of size 162$\times$54 and the output is a two-element vector that describes the probability of the hodogram being an FR or NFR. We use three convolution layers with kernel size 3 $\times$ 3, two of which are followed by max-pooling layers with kernel size 2 $\times$ 2. The convolution layer expands the dimension, whereas the max-pooling layers contract them. Finally, the array is flattened to 256 nodes and the network completes with two fully connected layers. These last layers take the dimensions from 256 inputs to 64 outputs, then 64 inputs to 2 outputs. All the convolution layers and the first fully connected layer use a Rectified Linear Units \citep[ReLU;][] {Nair2010RectifiedLU}) activation whereas the final layer uses a sigmoid activation function. The model and training are implemented with Tensorflow \citep{2016arXiv160304467A} version 2.10.0 in a Python 3.10.8 environment. Once the model is constructed, it is trained using realistic simulated FR and NFR hodograms. The preparation of the synthetic hodograms is described in the next section. \par
% In Figure \ref{}b, the DeepFRi model architecture is shown, where the grey rectangles show the output of each layers. 

The type of geoeffectiveness of a given FR is detected by M2. To classify the FR geoeffectiveness we consider the Dst index, which is an index of magnetic activity derived from a network of near-equatorial geomagnetic observatories that measures the decrease in the horizontal component of the Earth's field due to the ring current \citep{2001SSRv...98..343D}. We choose the Dst index threshold as $-50$ nT to include moderate to intense storms \citep{1994JGR....99.5771G}, in which the contribution of ICME FRs is prominent. We classify the events having $Dst_{min}<-50$ nT and $Dst_{min}>-50$ nT as `1' and `0', respectively. The M2 is a supervised binary-classifier machine learning model. We choose support vector machine (SVM) classifier -- a widely applied ML approach in solar and heliospheric physics and was first proposed by \citet{cortes1995support}. A general architecture of SVM model can be found in \citet{alvarez2021hybrid}. SVMs can identify both linear and nonlinear relationships between features by using linear and nonlinear kernels that can map these features into a higher dimensional space and thus SVMs may potentially explore previously unknown properties in the data. From previous studies (e.g., \citet{2003GeoRL..30.2039W, 2005JGRA..110.2213K}) it is well known that the duration of southward solar wind magnetic field component has a nonlinear relationship with the $Dst_{min}$, whereas the product of solar wind velocity and the minimum value of southward magnetic field intensity has a linear relationship with the $Dst_{min}$ \citep{2005GeoRL..3218103G, 2008JASTP..70.2078G}. Here, we use both linear and nonlinear kernels in M2. 
% The model is trained using $R_{Bz}$, $-B_{z_\mathrm{gsm},min}$, which is the minimum value of $-B_{zgse}$ during the real FR intervals, and the classes of their geoeffectiveness (class 0 for non-geoeffective and class 1 for geoeffective cases). 

\subsubsection{Preparation of training dataset} \label{Modeltrainingandevaluation}
\begin{figure}[!tbp]
  \centering
    \includegraphics[width=1\textwidth]{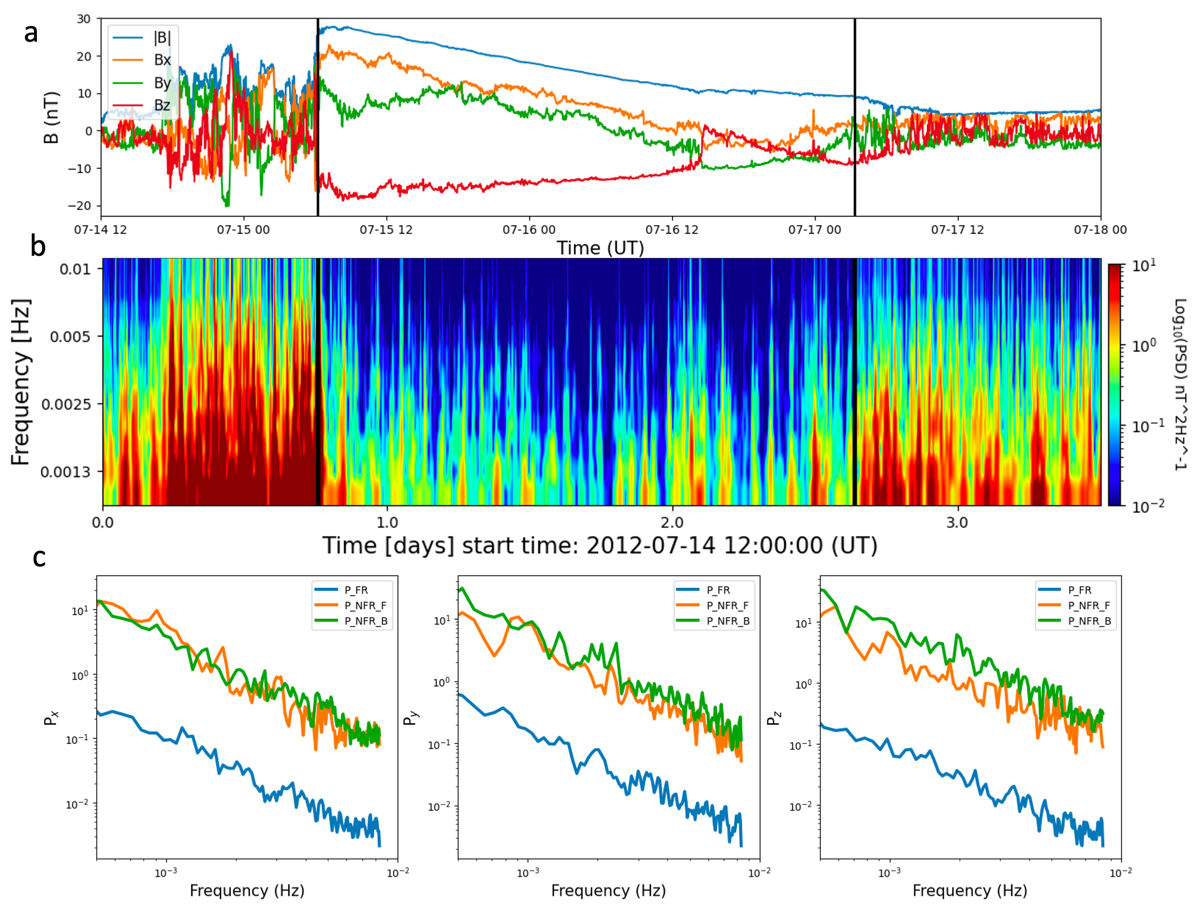}
  \caption{(a) Magnetic vector components ($B_x$, $B_y$, $B_z$) in GSE and magnetic field intensity ($|B|$) of solar wind containing an identifiable ICME FR bounded by vertical lines. (b) The wavelet PSD of $\mathbf{B}$ field fluctuations sampled at inertial range frequencies. (c) $P_x$, $P_y$, $P_z$ are the PSDs of $B_x$, $B_y$, $B_z$ field line fluctuations, respectively, where the field lines are normalized by maximum magnetic field intensity and belong to the ICME FR (blue), and non-FR segments (green and orange) before and after the ICME in 24-hours interval, respectively.   }
  \label{fig2}
\end{figure}

As the number of observed and cataloged events is not enough to train a neural network, in this work we generate several synthetic ICME FRs ($\mathbf{B_{FR}}(t)$) and NFRs ($\mathbf{B_{NFR}}(t)$) to train the model M1. The $\mathbf{B_{FR}}(t)$ is generated combining a physics-based FR model -- namely the circular-cylindrical \citep[CC;][]{2016ApJ...823...27N} model output $\mathbf{B_m}(t)$, with additive background noise $\delta \mathbf{B_{FR}}(t)$. Here, $\mathbf{B_m}(t)$ consists of the time series of each magnetic field component -- $B_{mx}(t), B_{my}(t), B_{mz}(t)$ obtained in the GSE coordinate system assuming several different spacecraft trajectories through the modeled CC flux ropes. The CC model provides a conventional smooth rotation in the FR magnetic field, while the additive noise $\delta B_{FR}(t)$ provides the type of fluctuations that are observed in real FRs. 
Our generation process for synthetic FR data is inspired by \citet{2020SoPh..295..131D} who used a modeled FR with additive Gaussian noise set called “5\% noise” and “10\% noise”, where the noise values were drawn from normal distributions having standard deviations of 0.05 and 0.1, respectively. The study did not implement physics-based fluctuations to their synthetic data. In our study, the background noise is modeled using the PSD of solar wind magnetic field fluctuations at inertial-range frequencies ($10^{-3} - 10^{-2}$ Hz). This frequency range corresponds to the spatial scales of magnetohydrodynamic turbulence. Typically, in this frequency range, magnetic field fluctuation power is higher in the solar wind than within FRs. This trend can be seen, for example, in the upstream wind and ICME FR of an event observed by Parker Solar Probe \citep[PSP;][]{2016SSRv..204....7F} at 0.25 au \citep{2020ApJ...900L..32G}. In Figure \ref{fig2}a, we show the magnetic field components along with the magnitude of an ICME FR and its upstream and downstream solar wind conditions observed by Wind at 1 au. In Figure \ref{fig2}b, we show the wavelet PSD of the magnetic field fluctuations sampled at inertial range frequencies. %Here, we use a 30-mins detrending factor, therefore, the y-axis ranges from $\sim5\times10^{-4} - 10^{-2}$. 
The vertical lines show the boundaries of the FR. \par

We use 399 ICME events from the \href{https://wind.nasa.gov/ICME_catalog/ICME_catalog_viewer.php}{Wind ICME catalog} during 1995--2021 to obtain the PSD of real ICME FRs and NFRs required to generate synthetic events. This catalog refers to the ICME FRs as magnetic obstacles (MOs) and notes their start and end boundaries as `MO start time' and `MO/ICME end time', respectively. The MOs having a sheath (a pile-up of solar wind commonly observed ahead of ICME FRs) are listed such that the sheath start time is listed as the `ICME start time'. The catalog also lists the minimum Dst value $Dst_{min}$ during each event and mean magnetic field intensity. We decided not to merge other catalog events to the \textit{Wind ICME catalog} because the criteria to select event boundaries are not exactly similar for all available catalogs.

\subsubsection*{Generation of synthetic flux ropes}
 Synthetic ICME FRs are generated by considering the local configuration of FRs as cylindrical structures with circular cross-sections. The model does not impose any condition on the internal force distribution but makes assumptions about the axial and poloidal current density components. In this work, we keep the ratio between the poloidal and axial current density components as $C1$ = 1, which implies that the structure can behave as force-free structure (see \citet{2016ApJ...823...27N} for more details). We generate several different unique FR magnetic field time series by varying the model inputs -- 
\begin{itemize}
    \item $\theta$ (the angle between the FR axis and ecliptic plane, -80$^\circ \le \theta \le 80^\circ$) and $\phi$ (an angle formed by FR axis and $+x_{gse}$, 5$^\circ \le \phi \le 355^\circ$) are chosen from equi-distributed points on surface of a sphere with polar angle $\theta$ and azimuthal angle $\phi$ \citep{deserno2004generate}.   
    
    % \item $\phi$ (an azimuth angle of the FR axis anti-clockwise positive from $+x_{gse}$) from 5$^\circ$ to 355$^\circ$ in steps of 5$^\circ$,
    \item $Y_0$ (the perpendicular distance between the spacecraft trajectory and FR center normalized to the FR radius) from 0 to 0.75 from either side of the FR center in steps of 0.05, and
    \item $H$ (handedness of the FR) either -1 or 1.
\end{itemize}   
This results in 161,000 simulated FR magnetic field vectors ($\mathbf{B_m}(t)$) normalized to the magnetic field intensity. \par 
To obtain the additive background profile $\delta B_{FR}(t)$ for generating realistic simulated FRs, we followed \citet{2021A&A...656A..13W, 2022JGRA..12730898W} that generated random and unique noise-field time series with PSD $P_{fr}(k)$ similar to the PSD of real FR magnetic fields. This method is inspired by an analogous process used to generate primordial
perturbations for dark matter N-body simulations \citep{2011MNRAS.415.2101H}. The $\delta \mathbf{B_{FR}}(t)$ is expressed as  
\begin{equation}
    \mathbf{\delta B_{FR}}(t)= \mathcal{F}^{-1} (\sqrt{P_\mathrm{fr}(k)} u(k)),
\end{equation}
where $u(t)$ is a random Gaussian sequence with mean 0 and standard deviation 1 and its Fourier transform $u(k)=\mathcal{F}(u(t))$. As Gaussian noise has a flat power spectrum, multiplying it in Fourier space with $\sqrt(P_{fr}(k))$ and transforming back to real space (time domain) provides a noisy sequence $\delta \mathbf{B_{FR}}(t)$ with PSD the same as $P_{fr}(k)$. We compute $P_{fr}(k)$ for each magnetic field component (normalized to magnetic field intensity) by applying Welch's algorithm \citep{1967ITAE...15...70W} on the real event intervals. By averaging several periodograms of overlapping segments that are shorter than the total length of the data, this approach averages out the noise in the data. The real event intervals are gathered from the \textit{Wind ICME catalog}. We use the `Welch' function available in the `signal' library of the `SciPy' package in Python, where $nperseg= 2^8$  is considered for the length of each overlapping segment at which the modified periodogram is computed. 
% As smaller segment length leads to better noise rejection, we choose the segment length 256 as $\sim$ 18\% of the average intervals of the real ICME magnetic obstacles at 1 au.
Before applying Welch's method, we detrended the normalized magnetic components with three different detrending time intervals $\Delta T$ -- [10, 30, 60] minutes. The detrending of signals helps remove any kind of pattern at the detrending timescale in the data. In Figure \ref{fig2}c, we show $P_{fr}(k)$ as a function of spacecraft-frame frequency for the ICME FR shown in \ref{fig2}a, with the PSD for each of the three FR magnetic field components in blue. The $\Delta T$ used to create $P_{fr}(k)$ in this case is 30 min. Therefore, the y-axis ranges from $\sim5\times10^{-4} - 10^{-2}$. The inertial-range spectral slope in the FR is found as $\sim-1.6$. The spectral index in the inertial range found in this case is consistent with various theories of Alfvénic MHD turbulence \citep{2022JPlPh..88e1501S}. \par

\begin{figure}[!tbp]
  \centering
    \includegraphics[width=1\textwidth]{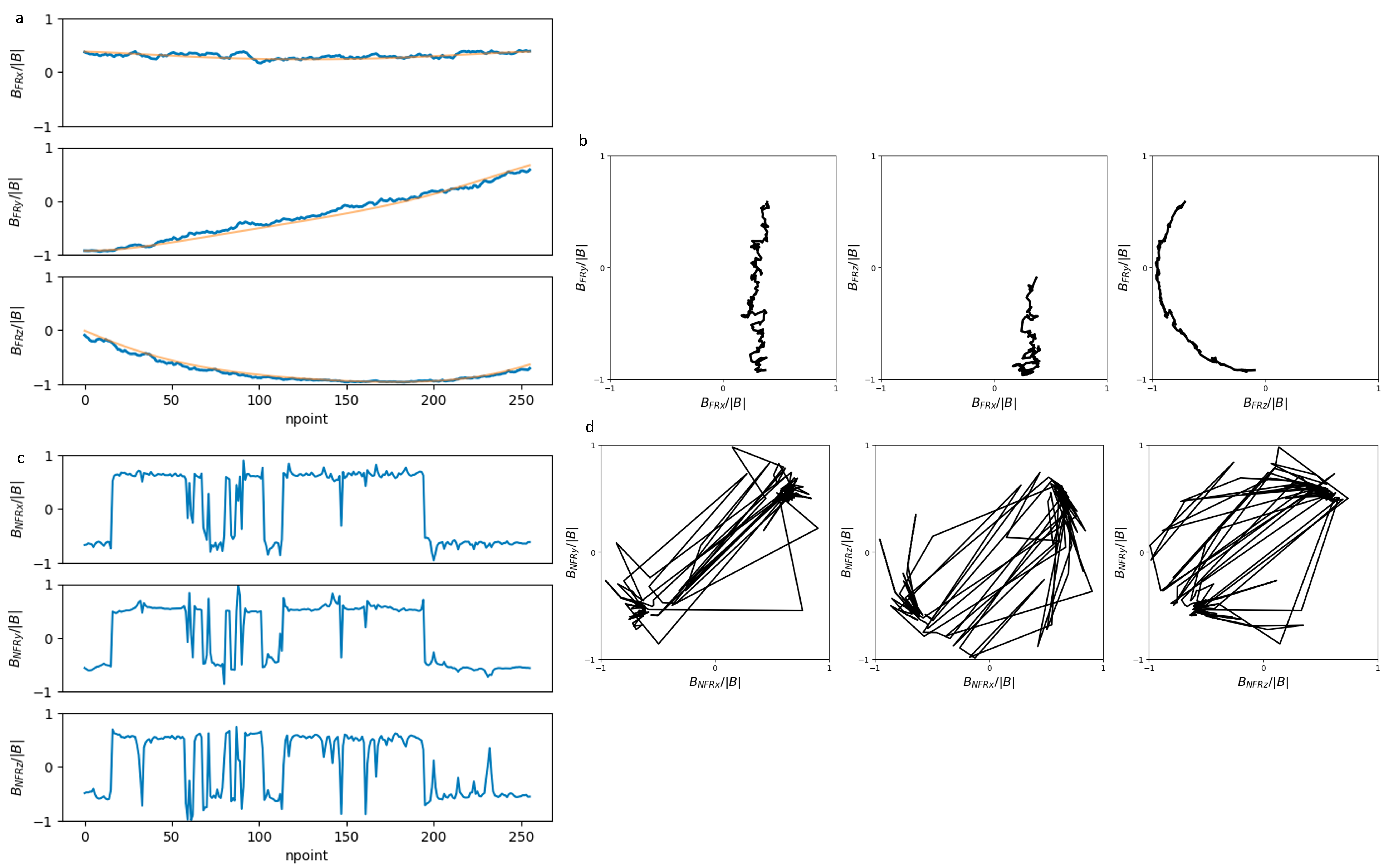}
  \caption{(a) Example of a simulated FR time series $\mathbf{B_{FR}}$ and (b) hodogram in $x_{gse}-y_{gse}$, $y_{gse}-z_{gse}$ and $z_{gse}-x_{gse}$ planes. The light-orange line over-plotted on (a) represents the $\mathbf{B_m}$ signal. (c) Example of a simulated NFR time series  $\mathbf{B_{NFR}}$ and its (d) hodogram in three different planes. The magnetic field vectors are normalized with intensity $|B|$. }
  \label{fig3}
\end{figure}

Using the \textit{Wind ICME catalog}, we generate 399 PSDs of real FR events, randomly pick a PSD to generate a unique $\delta B_{FR}(t)$ and add it to each component of $\mathbf{B_m}(t)$ to finally obtain a synthetic FR magnetic field $\mathbf{B_{FR}}(t)$, where 
\begin{equation}
    \mathbf{B_{FR}}(t) = \mathbf{B_m}(t)+\delta \mathbf{B_{FR}}(t)
\end{equation}
Note that here each of the $\mathbf{B_{FR}}(t)$ has $npoint$ number of points in it. Thus, we simulate a total number of 161,000 unique synthetic large-scale FR magnetic field time series with each having underlying PSDs equal to one of the real events. As this work focuses on the geometry of magnetic structures rather than the magnitude, we plotted each $\mathbf{B_{FR}}(t)$ in three different planes $x_{gse}-y_{gse}$, $y_{gse}-z_{gse}$ and $z_{gse}-x_{gse}$ in the -1 to 1 range and generated three different images of resolution 54 $\times$ 54, where each of them are called hodograms. Finally, the hodograms of three different planes are merged into a single image of size (Width) 162 $\times$ (Height) 54. In Figure \ref{fig3}a and b we show a simulated FR time series and hodogram, respectively, where the orange curves on Figure \ref{fig3}a represents $\mathbf{B_m}(t)$. The CC model inputs $\theta$, $\phi$, $Y_0$ and $H$ and the detrending factor $\Delta T$ for generating this particular synthetic FR is used as $60$ deg, $32$ deg, 0.65, $1$ and 30 min, respectively.   \par 

\subsubsection*{Generation of synthetic non-flux ropes}

Since we do not anticipate any typical rotation in NFRs, we only include the process of generating the solar wind background profile $\delta \mathbf{B_{NFR}}(t)$ to create synthetic NFRs. To create $\delta \mathbf{B_{NFR}}(t)$ we compute PSDs of the normalized magnetic field components of real NFRs, $P_{nfr}(k)$. From the Wind/MFI data, we randomly selected 2200 real unique NFR intervals having durations of 24 hours. A 24-hour duration for the NFRs was selected because this is equal to the average ICME duration at 1 au. Each of the NFRs starts at least 2 hours after the `ICME/MO end time' and ends at least 2 hours before `MO start time' mentioned in \textit{Wind ICME catalog}. Additionally, we obtain 303 sheath intervals from the\textit{Wind ICME catalog} which are bounded by `ICME start time' and `MO start time'. Thus we generate a total number of 2503 $P_{nfr}(k)$s, each of them corresponding to a real NFR that does not have any overlap with ICME MOs cataloged in the \textit{Wind ICME catalog}. In Figure \ref{fig2}c, we show with green and orange lines $P_{nfr}(k)$ as a function of spacecraft-frame frequency for the three different magnetic field components of two 24-hour NFR segments ahead of and after the ICME FR shown in Figure \ref{fig2}a. The spectral slope is found to be $\sim-1.8$ for both upstream and downstream NFR segments. The $\Delta T$ used to create $P_{nfr}(k)$ in this case, is 30 min. As no rotation in the NFRs' magnetic field components is expected, we generate the NFRs using Equation 2, where $P_{fr}$ is replaced by $P_{nfr}$. Thus,
\begin{equation}
    \mathbf{B_{NFR}}(t) = \delta \mathbf{B_{NFR}}(t) = \mathcal{F}^{-1} (\sqrt{P_\mathrm{nfr}(k)} u(k)).
\end{equation}
Once we generate a similar number of synthetic NFR time series as FRs, we obtain their hodograms in each plane and merge them into a single image of similar size as the synthetic FR hodograms. In Figure \ref{fig3}c and d, we show an example synthetic NFR time series and hodograms, respectively, where the $\Delta T$ for generating this particular synthetic NFR is used as 30 min. Finally, we generate nine sets of synthetic FR and NFR data, where each data set is prepared using a unique combination of $\Delta T$ values for both FRs and NFRs.

\subsubsection{Model training and evaluation}
Once we prepare the synthetic data set, we randomly select 30\% of the training data for validation. To train the network weights, we use the Adam optimizer \citep{Kingma2014AdamAM} with an initial learning rate of 0.001. We find that the accuracy and loss converge quickly for the validation set and infer that this may happen due to the simplicity of the classification problem in the simulation space. To avoid over-fitting we limit the training of the network to 50 epochs. Thus, we train nine M1 models of similar architecture using nine different sets of synthetic data. Using a system with a 10-core CPU processor and 16 GB RAM, we finish training each model in 58 minutes. After training each, we save their weights, evaluate each model with real cases, and derive the evaluation metrics. Finally, out of nine trained M1 models, we choose a model that provides the highest Area Under the Curve (AUC) value measuring the area under the `Receiver Operating Characteristic' (ROC) curve and designate this M1 as the DeepFRi model.  \par

For evaluating the performance of M1 models, we obtain 127 real ICME FRs having duration between 20-30 hours from \textit{Wind ICME catalog} and an equal number of non-ICME windows of 24 hours randomly chosen from Wind/MFI data. We use these events to evaluate the M1 models and derive evaluation metrics -- accuracy, precision, recall, F1 Score and AUC for each M1. These are the standard metrics for evaluating machine learning models. They are described using four terms, namely true positives (TP: cases where the predicted and original classification agree on being positive (i.e. FR)), true negatives (TN: cases where the predicted and original classification agree on being negative (i.e. NFR)), false positive (FP: if the prediction is positive (i.e. FR) but the ground truth is negative (i.e. NFR)), and false negative (FN: if the classification is negative (i.e. NFR) and the ground truth is positive (i.e. FR)). In Table \ref{t1}, column 1, we describe the evaluation metrics. \par 

\begin{figure}[!tbp]
  \centering
    \includegraphics[width=1\textwidth]{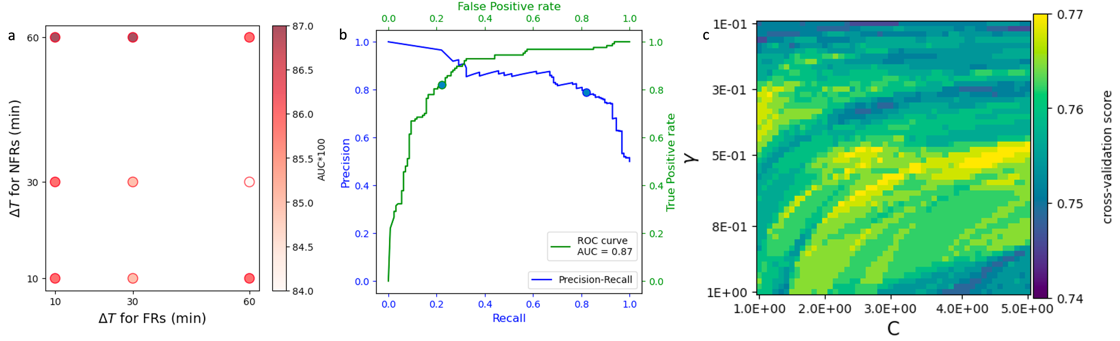}
  \caption{(a) The Area Under the Curve (AUC) values of nine M1 models with similar architecture trained with nine different combinations of simulated FR and NFR data. The detrending factors used to generate the simulated training data are 10, 30, and 60 mins. The maximum AUC value achieved as 0.87. (b) The ROC (green) and precision-recall (blue) curves corresponding to the M1 having maximum AUC accepted as the final model named DeepFRi. (c) Color map showing cross-validation scores obtained for different values of the hyperparameters $C$ [1-10$^{0.7}$] and $\gamma$ [10$^{-1}-1$] of the RBF kernel of SVM model in M2.  }
  \label{fig4}
\end{figure}

In Figure \ref{fig4}a, we display the AUC values of nine M1 models with circles of varying shades of red colour. We notice that training the same model architecture with different data sets prepared using different detrending factors leads to varying AUC values. The maximum obtained AUC is 0.87. In Figure \ref{fig4}b, we show the ROC (in green) and precision-recall (in blue) curves of one of the highest performing models named as DeepFRi trained with a set of synthetic FRs and NFRs, whose underlying PSDs are prepared using the linear detrending scales of $\Delta T = 10$ and $60$ minutes, respectively. The green and blue circles over-plotted to the curves corresponds to the decision threshold. 
% An hourly-scale linear detrending process can be better at capturing FR magnetic field fluctuations by more efficiently removing the impact of large-scale field line rotations. As NFRs do not have such rotations, a smaller detrending scale may perform better at capturing the fluctuations underlying the real NFR segments of the solar wind. 
The evaluation scores including accuracy, precision, recall and F1 Score of the DeepFRi, are indicated in Column 3 of Table \ref{t1}.  \par 

\begin{table}[t!]
\caption{Evaluation metrics of the DeepFRi and DicFR models. \label{tab:pms} \vspace*{-0.12in}}
\centering
\renewcommand{\arraystretch}{1.05}
\begin{tabular}{l@{\hskip .32in}c@{\hskip .22in}c@{\hskip .22in}c}
\toprule
 \textsc{\textbf{Performance Metrics}}& \textsc{\textbf{Description}}&\textsc{\textbf{DeepFRi performance}} & \textsc{\textbf{DicFR performance}} \\
\midrule
Accuracy & $\frac{TP+TN}{Total\  number\ instances}$&0.8 & 0.83 \\
Precision & $\frac{TP}{TP+FP}$& 0.78 & 0.85 \\
Recall & $\frac{TP}{TP+FN}$ &0.82 & 0.76 \\

F1-Score & $\frac{2 * Precision*Recall}{Precision+Recall}$& 0.8 & 0.805 \\
% AUC & & 0.87 & 0.83 \\

\bottomrule
\end{tabular}
\label{t1}
\end{table}

To train M2, we randomly select 75\% of ICMEs listed in the \textit{Wind ICME catalog}. %We employ the support vector machine (SVM) classifier which is a widely applied ML approach in solar and heliospheric physics and was first proposed by \citet{cortes1995support}. SVMs can identify both linear and nonlinear relationships between features by using linear and nonlinear kernels that can map these features into a higher dimensional space and thus SVMs may potentially explore previously unknown properties in the data. 
We use the Python scikit-learn library to implement SVM in M2. SVM has been integrated into the Python scikit-learn library \citep{pedregosa2011scikit}. It has open-source access and a well-established documentation \href{http://scikit-learn.org/stable/}{http://scikit-learn.org/stable/}. We use the stratified k-fold cross-validation method with k=5. It is implemented in the cross-validation module of the scikit-learn library. The non-linear radial basis function (RBF) kernel of SVM has two hyper-parameters -- ($`C'$ and $\gamma$) that need to be tuned to achieve the highest model performance. The $C$ is a regularization parameter that trades off correct classification versus maximizing the margin of the decision function. A low $C$ value corresponds to a large margin and simple decision function and a high $C$ value leads to overfitting. The $\gamma$ is a kernel width parameter that affects the shape of the class-dividing hyperplane. While using the RBF kernel, we preliminary set the range of $C$ and $\gamma$ as [$10^{-2},10^{2}$] \citep{2022ApJ...934..176P} and tune the parameters in that range to obtain the highest cross-validation score. Based on the performance, we further shrink the range of $C$ to [$10^{0},10^{0.7}$] and $\gamma$ to [$10^{-1},10^{0}$] and find the best-performing model with the highest cross-validation score of 0.77 at $C$ = 4.428, $\gamma$ = 0.543, and random state = 42. In Figure \ref{fig4}c, we provide a color-map showing the variation of cross-validation scores with different values of $C$ and $\gamma$. While using the linear kernel of SVM in M2, we find the cross-validation score to be 0.75. We compare our model performance with that of the K-nearest neighbor \citep[KNN;][]{Mucherino2009, 9065747} classifier. The KNN follows a similarity-based algorithm where the label of an event is assigned based on that of the majority of $K$ most similar examples. Here similarity is expressed in terms of the generalized concept of distance. The maximum cross-validation score achieved in the case of the KNN classifier is 0.75 with $K=12$. Therefore, we choose M2 with the SVM architecture and with a tuned RBF kernel to use as the DicFR model. \par   

To evaluate the performance of DicFR, we use the remaining 25\% cases from \textit{Wind ICME catalog}. In Column 4 of Table \ref{t1}, we provide the evaluation scores of the DicFR model.

\subsection{Post-processing}\label{Post-processing}
The DeepFRi model classifies each input window as FRs or NFRs and assigns them a binary value $V_{W} =1$ or 0, respectively. We first convert the hodograms to time-series data. Through a few post-processing steps, we derive a list of large-scale FR intervals containing the category of their geoeffectiveness. Below we describe the post-processing steps. \par 

% To prepare this catalog from the output of the CNN model, we follow a few steps. First, 
\textit{Step 1:} First we note the start time of each window that is predicted as $V_{W} =1$, by the model. If two consecutive windows (e.g. $W_1$ and $W_2$) with $V_{W} =1$ have start times are more than 24 hours apart, we record both the windows' end time and start time as corresponding to the previous FR's end and the next FR's start boundaries, respectively. Thus we check whether the same condition holds for the next two consecutive windows ($W_3$ and $W_4$), and so on. If the condition is not satisfied, we continue examining the rest of the window start times until the requirement is satisfied. Following this process, we generate a list (\textit{List 1$^\prime$}) where each row contains a start and end time of an interval and each data points inside the intervals belong to windows with $V_{W} =1$. \par
\begin{figure}[!tbp]
  \centering
    \includegraphics[width=0.7\textwidth]{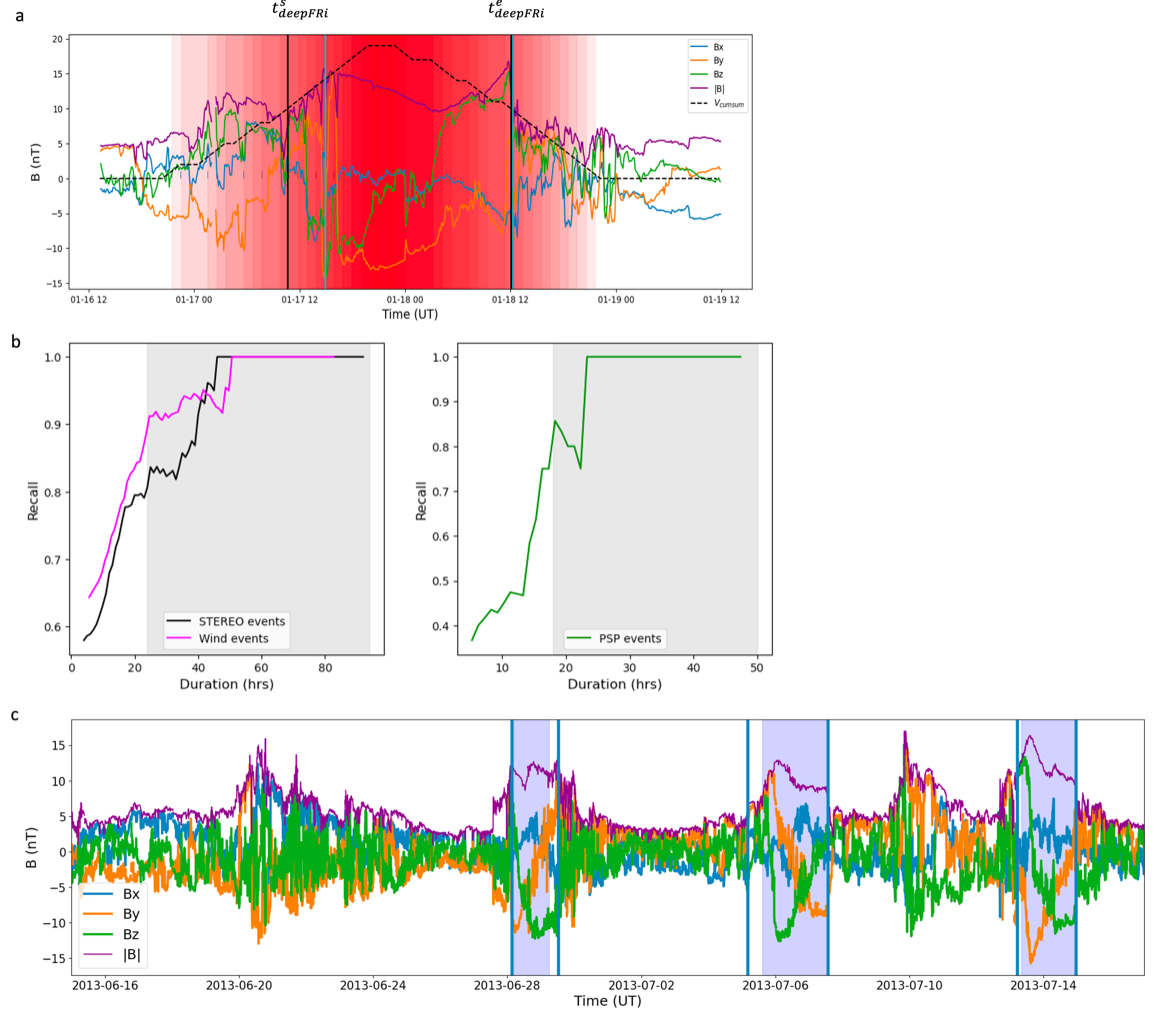}
  \caption{(a) A solar wind magnetic field interval during 2013/01/16 12:00 - 2013/01/19 12:00 UT. Each red-shaded rectangle has duration of 24 hours and $V_W=1$. The black dashed line indicates $V_{cumsum}$. The black vertical lines mark the $t^{s}_{DeepFRi}$ and $t^{e}_{DeepFRi}$ interval. The cyan vertical lines show boundaries of the event obtained from the \textit{Wind ICME catalog}. (b) Sensitivity of the DeepFRi model as a function of ICME FR duration. The purple, black and green colored lines correspond to the ICMEs observed by Wind, STEREO and PSP spacecraft, respectively. The recall values belonging to the gray-shaded regions are $\ge 0.8$. (c) A 30-day interval of solar wind magnetic field vectors where the blue shaded regions indicate auto-detected geoeffective FR intervals. The solid vertical lines indicate FR start and end times obtained from RL. }
  \label{fig5}
\end{figure}

 
\textit{Step 2:} In the next step, we compute the cumulative sum $V_{cumsum}$ of $V_{W}$ values of the windows within each interval in \textit{List 1$^\prime$}. Note that each interval may contain a single or multiple number of windows. Intervals containing multiple windows can have $V_{cumsum}$ values varying over time. Once we derive the $V_{cumsum}$ for each interval, we obtain its full width at half maximum (FWHM) duration and define it as an auto-detected FR interval. These intervals are noted in \textit{List 1}. In Figure \ref{fig5}a, we show an interval containing solar wind magnetic field vectors which have been used as input to our model. The red-shaded rectangles indicate the times when $V_{W} =1$. Using a black dashed curve, we show $V_{cumsum}$. We note the start and end times of the FR interval as $t^{s}_{DeepFRi}$ and $t^{e}_{DeepFRi}$ in black vertical lines, respectively. The cyan vertical lines show the FR boundaries noted in the \textit{Wind ICME catalog}. \par

\textit{Step 3:} Finally, we use a magnetic field magnitude threshold $B_{Th}$ and check whether each interval's mean magnetic field intensity $B_{mean}$ crosses $B_{Th}$. To obtain $B_{Th}$, we derive the cumulative distribution function ($F(X)=P[X\le x]$), where X is the $B_{mean}$ of real events from \textit{Wind ICME catalog} with duration 24 hours or more. We assign $B_{Th}$ such as $P(X \ge B_{Th})=0.9$. Once the auto-detected FR intervals satisfy the threshold criteria, they are put in the list \textit{List 2}.\par

\textit{Step 4:}  We derive the geoeffective properties of the FR intervals required to provide input to the DicFR model. The DicFR model classifies the intervals as class 1 (geoeffective) or class 0 (non-geoeffective) and a \textit{Final List} is created that contains $t^{s}_{DeepFRi}$, $t^{e}_{DeepFRi}$ and FR geoeffectiveness category (class 0 or 1). 

\section{Results and Discussion}\label{sec:3}

We now assess the performance of our pipeline by comparing its output to the existing manually detected open-source ICME lists. For this purpose, we use the list of ICME events observed by STEREO A and B during 2007--2020 available at \textit{STEREO ICME catalog} and the reference list (RL) used by \citet{2019ApJ...874..145N}. The RL is a union of different Wind ICME lists obtained from \citet{2006SoPh..239..393J, 2006AnGeo..24..215L, 2010SoPh..264..189R, 2016SoPh..291.2419C} and \citet{2018SoPh..293...25N} and compiled by \citet{2019ApJ...874..145N} available at \href{https://github.com/gautiernguyen/Automatic-detection-of-ICMEs-at-1-AU-a-deep-learning-approach}{https://github.com/gautiernguyen/Automatic-detection-of-ICMEs-at-1-AU-a-deep-learning-approach} and a recently created list of ICMEs observed by PSP during 2018 - 2022 \citep{2024ApJ...966..118S}. For this purpose, we use magnetic field data derived from the In-Situ Measurements of Particles and CME Transients \citep[IMPACT;][]{2008SSRv..136..117L} instrument onboard STEREO twin spaceceraft and Fields \citep[FIELDS;][]{2016SSRv..204...49B} investigation of PSP. It is important to note that any existing ICME FR catalogs are likely not exhaustive and the event start and end times are based on the interpretation of data by human expertise. We first checked how our model identifies the listed events in the catalogs. We then chose a seven-year interval (2008--2014) of continuous Wind/MFI data, prepared the \textit{Final List}, compared that with the open-source lists, and obtained the recall and precision values for both models in the pipeline. In Figure \ref{fig5}b, we show the sensitivity of DeepFRi as a function of duration of ICME FRs obtained from different spacecraft. In the left panel, we show how the model's recall value changes while testing it on real events with duration ranging from $\sim$5-90 hours observed at 1 au. The gray-shaded region corresponds to recall $\ge 0.8$. The high recall values in the shaded region indicate the better performance of DeepFRi while testing it with large-scale FRs. We notice that the recall becomes 0.886 while testing the model with events having $\ge$ 24 hours interval  obtained from RL list. However, we observe that even if we used 24-hour window in training purpose, our model identified 75\% of the ICME FRs observed by Wind having duration $\ge$15 hours. \par 

We provided 2008-2014 Wind/MFI data to our pipeline. The data is first pre-processed and provided to the DeepFRi model following the steps described in Section \ref{pre-processing}. The DeepFRi model classified 639 windows as class 1, where each window lasts $\sim$24 hours and two consecutive window start times are minimum 1 hour apart. By following \textit{Step 1 -- Step 3} described in Section \ref{Post-processing} we generated a \textit{List 1} with 296 FRs and \textit{List 2} with 110 FR intervals having mean intensity greater than $B_{Th}$. In Figure \ref{fig5}c, we show a small 30-day subset of the whole interval 2008 - 2014. The blue-shaded regions show the intervals identified by our pipeline as geoeffective FRs presented in the \textit{Final List}. The vertical solid lines are FR boundaries noted in RL. 

\par 
We compared our auto-generated FR lists to the RL events and obtained the number of TPs, FPs, TNs, and FNs. We derived model precision and recall following the process described in Table \ref{t1}. Out of 64 events with duration $\ge24$ hours obtained from RL during 2008-2014, we found 56 events that are detected by our pipeline and presented in \textit{List 1}. We computed the recall as 0.875. while comparing \textit{List 2} to the RL events, we find the recall as 0.8 and precision as 0.56. The high number of FPs suggests that the DeepFRi model identifies a number of rotations in solar wind magnetic field that may not correspond to ICME FRs. Including additional solar wind features e.g. temperature, proton $\beta$, velocity etc in detection process may help reducing the FPs. 
\par

\begin{figure}[!tbp]
  \centering
    \includegraphics[width=0.8\textwidth]{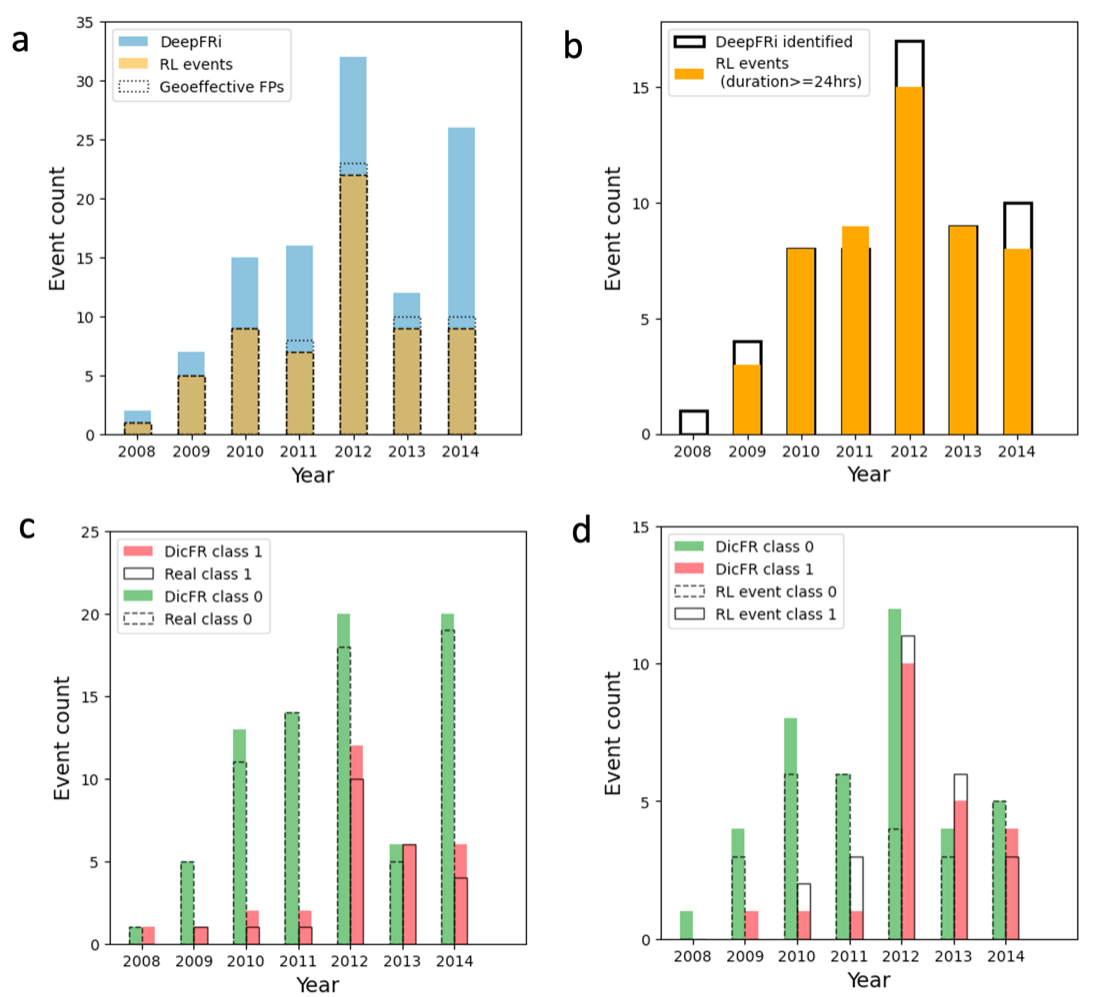}
  \caption{(a) Histogram with blue bars showing the number of events identified by the DeepFRi model during 2008-2014. The listed number of ICMEs detected by DeepFRi is shown using gold bars. The false positive events that have $Dst_{min}<-50 $nT are represented by the region between dotted and dashed lines at the bars of 2011--2014. (b) The number of ICMEs listed in RL each year between 2008 and 2014 with duration $\ge24$ hours are shown with orange bars. The solid line bars represent DeepFRi-identified ICMEs belonging to the RL. (c) DicFR model-classified (class 1 (red): geoeffective and class 0 (green): non-geoeffective) FRs obtained from DeepFRi each year between 2008-2014. The solid and dashed-line bars represent the real geoeffective and non-geoeffective classes found from the Dst index available at the \href{https://omniweb.gsfc.nasa.gov/form/omni_min.html}{OMNIWeb database.} (d) Solid and dashed line bars represent class 1 and class 0 events listed in the RL. Green and red bars represent DicFR-identified classes of the events.   }
  \label{fig6}
\end{figure}

In Figure \ref{fig6}a and b we summarise the result that we found while applying the DeepFRi model to detect ICME FRs during the 2008 - 2014 time period. The blue bars in Figure \ref{fig6}a indicate the number of FRs detected by DeepFRi each year. The over-plotted gold bars show the number of events that belong to the RL and are also detected by our DeepFRi model. The number of events listed in RL for each year between 2008 and 2014 with duration more than or equal to 24 hours is displayed in Figure \ref{fig6}b by the orange bars. Black bars are superimposed on the orange ones to display the events listed in RL detected by DeepFRi.

\par

\citet{2019ApJ...874..145N} used 33 different magnetic field and plasma parameters directly measured and derived from in situ observations to automatically detect ICME FRs using a CNN architecture. During 2010 – 2015, using Wind spacecraft data, they obtained a maximum recall of 0.84 and average precision of 0.697. When they considered only the magnetic field magnitude and component data, they found an average precision of 0.593 that is similar to what we obtain. Using Deep Residual U-Net model and training with the same 33 FR parameters used by \citet{2019ApJ...874..145N}, a study by \citet{ 2022SpWea..2003149R} found the recall and precision to be $0.67$ and $0.7$, respectively. Our pipeline finds the start of an ICME present in the \textit{Wind ICME catalog} with a mean absolute error (MAE) of $\sim7$ hours, and the end time with an MAE of $\sim6$ hours. In both \citet{2019ApJ...874..145N} and \citet{2022SpWea..2003149R}, the MAE values for the ICME start and end are less than that of our result because while training, they used a more sophisticated set of features and multiple sliding window sizes. Moreover, they optimized the decision threshold such that their model can determine the ICME intervals with high precision.
\par

With careful manual observation of solar wind plasma and magnetic data, \citet{2024ApJ...966..118S} identified ICMEs observed by PSP during 2018 - 2022 at heliocentric distances in the range of 0.23 - 0.83 au. We notice that the DeepFRi model can detect more than 80\% of listed PSP ICMEs when their duration is $\ge$ 18 hours. In the right panel of Figure \ref{fig5}b, we show the change of recall values while the model performance is tested on PSP events whose duration ranges from $\sim$6-48 hours. Note that the number of FNs increases because we train our model using 24 hours magnetic field data window and the PSD of fluctuations at 1 au. \citet{2023ApJ...956L..30G} studied the turbulence properties of ICMEs in the inner heliosphere and found that with increasing heliocentric distance the power of magnetic field fluctuation decreases. Using PSP, \citet{2020ApJS..246...53C} showed the variation of solar wind magnetic field PSD at different heliocentric distances. They found a significant radial evolution of the power of magnetic field fluctuations and the inertial range spectral slope of magnetic field PSDs. Therefore, for achieving better results at distances less than 1 au using our model, the preparation of the training dataset should include magnetic field PSDs at those distances. 

\par
Now, we compare the geoeffectiveness of auto-detected FRs obtained from the pipeline and available in the \textit{Final List} with the real minimum \textit{Dst} index during these intervals and RL events. During 2008 - 2014, we found 110 auto-detected FRs with the boundaries identified by $t^{s}_{DeepFRi}$ and $t^{e}_{DeepFRi}$ using our pipeline. We derive $R_{Bz}$ and $-B_{z_\mathrm{gsm},min}$ for each interval and provide these as inputs to the DicFR model. The model provides a binary output that indicates if the interval contains a $Dst_{min}<-50$ nT or not. Once the geoeffectiveness is predicted, the \textit{Final List} is generated following \textit{Step 4} described in Section \ref{Post-processing}. In our local system, it took $\sim$30 minutes to generate the \textit{Final List} from the 2008 - 2014 interval of raw Wind/MFI data. After comparing with the geoeffectiveness category obtained from real $Dst_{min}$ values of the auto-detected FR intervals, we find the accuracy of the model to be 0.88, recall to be 0.8, precision to be 0.77 and F1-Score to be 0.78. The \textit{Final List} may include non-ICME FRs, whereas the DicFR model is trained only on ICME FRs. Furthermore, certain events might have $R_{Bz}$ and $-B_{z_\mathrm{gsm},min}$ parameter values similar to those of geoeffective FRs, but their pressure and velocity might not be similar to those of geoeffective ones. This may lead to the generation of more false positive events. \par

In Figure \ref{fig6}c and d we summarise the DicFR model output using histograms. The green and red bars in the plots indicate the number of DeepFRi-detected FR intervals classified as non-geoeffective (class 0) and geoeffective (class 1), respectively. In Figure \ref{fig6}c, the solid and dashed-line bars superimposed on the red and green bars indicate the number of DeepFRi detected events that are geoeffective and non-geoeffective in reality, respectively. In Figure \ref{fig6}d, the black solid and dashed-line bars indicate the number of geoeffective and non-geoeffective events as identified in the RL. The red and green bars show the number of events correctly classified by DicFR.

\section{Summary and Conclusion}\label{sec:4}
This work provides a procedure to automatically detect heliospheric large-scale flux ropes (FRs) in solar wind data obtained from spacecraft at 1 au and classify their geoeffectiveness. We establish a pipeline that accepts solar wind magnetic field data from spacecraft and produces a list that contains auto-detected FR boundaries and information on whether the FR interval can produce moderate to intense geomagnetic storms or not. To accomplish this operation, the pipeline uses two models called DeepFRi and DicFR which are based on two ML techniques -- convolutional neural network (CNN) and support vector machine (SVM), respectively. The DeepFRi model is trained with simulated FR (positive class) and non-FR (negative class) events prepared by combining a modeled magnetic topology of FRs and power spectral densities of the magnetic field fluctuations in the solar wind. The DicFR model is trained with two geoeffective characteristics -- maximum southward magnetic field intensity and duration of southward magnetic field with respect to the whole FR intervals. Furthermore, these two models only use solar wind magnetic field parameters. \par
% Despite being trained with ICME-related FRs, DeepFRi can detect non-ICME-related FRs like small transients and small-scale flux ropes in the solar wind. This happens because these structures share the same magnetic topology with ICME FRs. Out of 399 listed ICME events in the \textit{Wind ICME} catalog observed during 1995 - 2021, DeepFRi can detect 338 events correctly. During the testing period (2008 - 2014), the model generates 82 false positive events and correctly identifies 88 out of 108 events from the \textit{Wind ICME} catalog. While testing the DicFR model using parameters of FRs detected by DeepFRi, it outputs 9 ($\sim20\%$) false negatives out of 44 events and 19 ($\sim10\%$) false positives out of 180 events.   \par
The pipeline we propose here introduces a way of automatically now-casting FRs including their geoeffective characteristics. The pipeline can be used for space weather operations with further improvements. Currently, the pipeline uses calibrated and processed magnetic field data. To utilise this in auto-flagging space weather events in real-time and issuing alert, additional data pre-processing steps are required. This includes the preparation and calibration of raw data obtained straight from instruments. We use the existing knowledge of FRs' generic structure, its duration, and FR and NFR magnetic field fluctuations properties at 1 au to train the models in our pipeline. In this work, we mainly focus on detection of 1-au ICME FRs and their geoeffectiveness. While applying this pipeline at distances less than 1 au, it under-performs, which reflects the changing FR duration and solar wind fluctuations with radial distance from the Sun. Therefore, before using this pipeline at other radial distances, re-training of the models based on the local FR and NFR properties is required.  \par

The reason for considering only magnetic field information in this work is mainly the lower availability of continuous solar wind plasma and composition information from probes at different locations. For example, during the cruise phase, a planetary mission Juno spent 5 years in the solar wind between 1 - 5.4 au and allowed us to study only the magnetic properties of ICMEs \citep{2021ApJ...923..136D} using its Magnetic Field Investigation instrument \citep[MAG;][]{2017SSRv..213...39C}. Another planetary mission, BepiColombo \citep{2021SSRv..217...90B}, which aims to understand Mercury, provides magnetic data but no solar wind plasma information. Therefore, our framework is capable of detecting FRs from spacecraft that provide only magnetic data, and also from Solar Orbiter \citep{2020A&A...642A...1M} and PSP, even when one or several plasma measuring instruments malfunction and are unable to provide data coverage. It can also be applicable to other spacecraft like Cluster, Helios, Ulysses, Venus Express, etc. that have measured the solar wind magnetic field over the past two decades throughout the heliosphere. However, training with only the magnetic field features leaves us with a number of false positive cases. While our pipeline may be helpful for space weather event identification purposes, we recommend manually verifying each auto-detected event utilizing its plasma and composition properties for event-based scientific study. After examining the false negative events, we have determined that our pipeline is unable to identify events that are too close to each other to be distinguished separately, have a short duration, or lack any rotation. However, an ensemble of training dataset with different windows may lead to success in the identification of short-duration ICME FRs and thus may decrease the number of false negative events. In future, we intend to modify the models in our pipeline to facilitate the identification of short-duration ICMEs.      \par 
Due to the lack of a very large FR database, we use synthetic data to train our model based on the architecture of a handwriting recognition model. For preparing the training dataset, we utilize a simplistic physics-based FR model that is unable to provide complicated magnetic configurations of distorted FRs. That may lead to the misclassification of solar wind data. This means that creating intricate FR models with more observable features is necessary, as is incorporating the models to produce better training data. Recently a study by \citet{2023ApJ...947...79N} has established a technique that may enable the modeling of realistic distortion of FR cross-sections. This work only aims to classify solar wind in flux rope and non-flux rope segments. To make our framework more comprehensive, we intend in the future to automatically classify the detected FRs in different categories based on their rotation, distortion and more complex structures. This may call for more advanced and complicated FR models to prepare training dataset.\par 

The performance of the second model, DicFR, which detects the geoeffectiveness of the DeepFRi-identified FRs in the pipeline, could be improved by including parameters like solar wind velocity and proton density during the FR interval. As we focused only on magnetic field information, we decided not to involve any plasma parameters in this particular study. We employed ICME-related FRs to train both the DeepFRi and DicFR models since they are more likely to cause geomagnetic storms than non-ICME ones. Integrating the second model into our pipeline makes it more favorable to space weather applications. Our tool could substantially contribute to space weather now-casting and forecasting processes.

\par
The trained ML model weights, codes that are used to apply the pipeline to the solar wind data and \textit{List 1}, \textit{List 2} and \textit{Final List} derived during the test period 2008 - 2014 are made available in \href{https://github.com/spal4532/DeepFRi-and-DicFR}{https://github.com/spal4532/DeepFRi-and-DicFR}. Users may download our pipeline in their local system and run on real 1-au solar wind data to prepare auto-detected FR lists and their geoeffectiveness.\par 
\hfill \break
% \begin{acknowledgments}
% \nolinenumbers
We acknowledge the data source \href{https://cdaweb.gsfc.nasa.gov/istp_public/}{Goddard Space Flight Center's Space Physics Data Facility (SPDF)} from where we downloaded the solar wind data and the open-source catalogs of \href{https://wind.nasa.gov/ICME_catalog/ICME_catalog_viewer.php}{Wind ICME}, \href{https://stereo-ssc.nascom.nasa.gov/data/ins_data/impact/level3/}{STEREO observed ICMEs} and reference ICME list from \href{https://github.com/gautiernguyen/Automatic-detection-of-ICMEs-at-1-AU-a-deep-learning-approach}{https://github.com/gautiernguyen/Automatic-detection-of-ICMEs-at-1-AU-a-deep-learning-approach}. S.P and A.J.W are thankful for the support of NASA Postdoctoral Program fellowship, L.K.J. thanks for the support of the STEREO mission and Heliophysics Guest Investigator Grant 80NSSC23K0447. T.N-C. thanks for the support of the Solar Orbiter and Parker Solar Probe missions, Heliophysics Guest Investigator Grant 80NSSC23K0447 and the GSFC-Heliophysics Innovation Funds. S.W.G. is supported by the Research Council of Finland (grants 338486 and 346612). 

% \end{acknowledgments}

\bibliographystyle{aasjournal}
\bibliography{bibliography}

\begin{thebibliography}{}
\expandafter\ifx\csname natexlab\endcsname\relax\def\natexlab#1{#1}\fi
\providecommand{\url}[1]{\href{#1}{#1}}
\providecommand{\dodoi}[1]{doi:~\href{http://doi.org/#1}{\nolinkurl{#1}}}
\providecommand{\doeprint}[1]{\href{http://ascl.net/#1}{\nolinkurl{http://ascl.net/#1}}}
\providecommand{\doarXiv}[1]{\href{https://arxiv.org/abs/#1}{\nolinkurl{https://arxiv.org/abs/#1}}}

\bibitem[{{Abadi} {et~al.}(2016){Abadi}, {Agarwal}, {Barham}, {Brevdo}, {Chen}, {Citro}, {Corrado}, {Davis}, {Dean}, {Devin}, {Ghemawat}, {Goodfellow}, {Harp}, {Irving}, {Isard}, {Jia}, {Jozefowicz}, {Kaiser}, {Kudlur}, {Levenberg}, {Mane}, {Monga}, {Moore}, {Murray}, {Olah}, {Schuster}, {Shlens}, {Steiner}, {Sutskever}, {Talwar}, {Tucker}, {Vanhoucke}, {Vasudevan}, {Viegas}, {Vinyals}, {Warden}, {Wattenberg}, {Wicke}, {Yu}, \& {Zheng}}]{2016arXiv160304467A}
{Abadi}, M., {Agarwal}, A., {Barham}, P., {et~al.} 2016, arXiv e-prints, arXiv:1603.04467, \dodoi{10.48550/arXiv.1603.04467}

\bibitem[{{\'A}lvarez-Alvarado {et~al.}(2021){\'A}lvarez-Alvarado, R{\'\i}os-Moreno, Obreg{\'o}n-Biosca, Ronquillo-Lomel{\'\i}, Ventura-Ramos~Jr, \& Trejo-Perea}]{alvarez2021hybrid}
{\'A}lvarez-Alvarado, J.~M., R{\'\i}os-Moreno, J.~G., Obreg{\'o}n-Biosca, S.~A., {et~al.} 2021, Applied Sciences, 11, 1044

\bibitem[{{Asensio Ramos} {et~al.}(2023){Asensio Ramos}, {Cheung}, {Chifu}, \& {Gafeira}}]{2023LRSP...20....4A}
{Asensio Ramos}, A., {Cheung}, M. C.~M., {Chifu}, I., \& {Gafeira}, R. 2023, Living Reviews in Solar Physics, 20, 4, \dodoi{10.1007/s41116-023-00038-x}

\bibitem[{{Bale} {et~al.}(2016){Bale}, {Goetz}, {Harvey}, {Turin}, {Bonnell}, {Dudok de Wit}, {Ergun}, {MacDowall}, {Pulupa}, {Andre}, {Bolton}, {Bougeret}, {Bowen}, {Burgess}, {Cattell}, {Chandran}, {Chaston}, {Chen}, {Choi}, {Connerney}, {Cranmer}, {Diaz-Aguado}, {Donakowski}, {Drake}, {Farrell}, {Fergeau}, {Fermin}, {Fischer}, {Fox}, {Glaser}, {Goldstein}, {Gordon}, {Hanson}, {Harris}, {Hayes}, {Hinze}, {Hollweg}, {Horbury}, {Howard}, {Hoxie}, {Jannet}, {Karlsson}, {Kasper}, {Kellogg}, {Kien}, {Klimchuk}, {Krasnoselskikh}, {Krucker}, {Lynch}, {Maksimovic}, {Malaspina}, {Marker}, {Martin}, {Martinez-Oliveros}, {McCauley}, {McComas}, {McDonald}, {Meyer-Vernet}, {Moncuquet}, {Monson}, {Mozer}, {Murphy}, {Odom}, {Oliverson}, {Olson}, {Parker}, {Pankow}, {Phan}, {Quataert}, {Quinn}, {Ruplin}, {Salem}, {Seitz}, {Sheppard}, {Siy}, {Stevens}, {Summers}, {Szabo}, {Timofeeva}, {Vaivads}, {Velli}, {Yehle}, {Werthimer}, \& {Wygant}}]{2016SSRv..204...49B}
{Bale}, S.~D., {Goetz}, K., {Harvey}, P.~R., {et~al.} 2016, \ssr, 204, 49, \dodoi{10.1007/s11214-016-0244-5}

\bibitem[{{Baruah} {et~al.}(2024){Baruah}, {Roy}, {Sinha}, {Palmerio}, {Pal}, {Oliveira}, \& {Nandy}}]{2024SpWea..2203716B}
{Baruah}, Y., {Roy}, S., {Sinha}, S., {et~al.} 2024, Space Weather, 22, e2023SW003716, \dodoi{10.1029/2023SW003716}

\bibitem[{{Benkhoff} {et~al.}(2021){Benkhoff}, {Murakami}, {Baumjohann}, {Besse}, {Bunce}, {Casale}, {Cremonese}, {Glassmeier}, {Hayakawa}, {Heyner}, {Hiesinger}, {Huovelin}, {Hussmann}, {Iafolla}, {Iess}, {Kasaba}, {Kobayashi}, {Milillo}, {Mitrofanov}, {Montagnon}, {Novara}, {Orsini}, {Quemerais}, {Reininghaus}, {Saito}, {Santoli}, {Stramaccioni}, {Sutherland}, {Thomas}, {Yoshikawa}, \& {Zender}}]{2021SSRv..217...90B}
{Benkhoff}, J., {Murakami}, G., {Baumjohann}, W., {et~al.} 2021, \ssr, 217, 90, \dodoi{10.1007/s11214-021-00861-4}

\bibitem[{{Bothmer} \& {Schwenn}(1998)}]{1998AnGeo..16....1B}
{Bothmer}, V., \& {Schwenn}, R. 1998, Annales Geophysicae, 16, 1, \dodoi{10.1007/s00585-997-0001-x}

\bibitem[{{Burlaga} {et~al.}(1981){Burlaga}, {Sittler}, {Mariani}, \& {Schwenn}}]{1981JGR....86.6673B}
{Burlaga}, L., {Sittler}, E., {Mariani}, F., \& {Schwenn}, R. 1981, \jgr, 86, 6673, \dodoi{10.1029/JA086iA08p06673}

\bibitem[{{Cane} {et~al.}(2000){Cane}, {Richardson}, \& {St. Cyr}}]{2000GeoRL..27.3591C}
{Cane}, H.~V., {Richardson}, I.~G., \& {St. Cyr}, O.~C. 2000, \grl, 27, 3591, \dodoi{10.1029/2000GL000111}

\bibitem[{{Cartwright} \& {Moldwin}(2008)}]{2008JGRA..113.9105C}
{Cartwright}, M.~L., \& {Moldwin}, M.~B. 2008, Journal of Geophysical Research (Space Physics), 113, A09105, \dodoi{10.1029/2008JA013389}

\bibitem[{{Cartwright} \& {Moldwin}(2010)}]{2010JGRA..115.8102C}
---. 2010, Journal of Geophysical Research (Space Physics), 115, A08102, \dodoi{10.1029/2009JA014271}

\bibitem[{{Chandorkar} {et~al.}(2017){Chandorkar}, {Camporeale}, \& {Wing}}]{2017SpWea..15.1004C}
{Chandorkar}, M., {Camporeale}, E., \& {Wing}, S. 2017, Space Weather, 15, 1004, \dodoi{10.1002/2017SW001627}

\bibitem[{{Chen} {et~al.}(2020){Chen}, {Bale}, {Bonnell}, {Borovikov}, {Bowen}, {Burgess}, {Case}, {Chandran}, {de Wit}, {Goetz}, {Harvey}, {Kasper}, {Klein}, {Korreck}, {Larson}, {Livi}, {MacDowall}, {Malaspina}, {Mallet}, {McManus}, {Moncuquet}, {Pulupa}, {Stevens}, \& {Whittlesey}}]{2020ApJS..246...53C}
{Chen}, C.~H.~K., {Bale}, S.~D., {Bonnell}, J.~W., {et~al.} 2020, \apjs, 246, 53, \dodoi{10.3847/1538-4365/ab60a3}

\bibitem[{{Chen} {et~al.}(2022){Chen}, {Deng}, {Li}, {Li}, {Chen}, {Chen}, \& {Luo}}]{2022ApJS..259....8C}
{Chen}, J., {Deng}, H., {Li}, S., {et~al.} 2022, \apjs, 259, 8, \dodoi{10.3847/1538-4365/ac4587}

\bibitem[{{Chen} \& {Hu}(2020)}]{2020ApJ...894...25C}
{Chen}, Y., \& {Hu}, Q. 2020, \apj, 894, 25, \dodoi{10.3847/1538-4357/ab8294}

\bibitem[{{Chen} \& {Hu}(2022)}]{2022ApJ...924...43C}
---. 2022, \apj, 924, 43, \dodoi{10.3847/1538-4357/ac3487}

\bibitem[{{Chen} {et~al.}(2019){Chen}, {Hu}, \& {le Roux}}]{2019ApJ...881...58C}
{Chen}, Y., {Hu}, Q., \& {le Roux}, J.~A. 2019, \apj, 881, 58, \dodoi{10.3847/1538-4357/ab2ccf}

\bibitem[{{Chi} {et~al.}(2016){Chi}, {Shen}, {Wang}, {Xu}, {Ye}, \& {Wang}}]{2016SoPh..291.2419C}
{Chi}, Y., {Shen}, C., {Wang}, Y., {et~al.} 2016, \solphys, 291, 2419, \dodoi{10.1007/s11207-016-0971-5}

\bibitem[{{Choi} {et~al.}(2012){Choi}, {Moon}, {Vien}, \& {Park}}]{2012JKAS...45...31C}
{Choi}, S.-H., {Moon}, Y.-J., {Vien}, N.~A., \& {Park}, Y.-D. 2012, Journal of Korean Astronomical Society, 45, 31, \dodoi{10.5303/JKAS.2012.45.2.31}

\bibitem[{Cire\c{s}an {et~al.}(2011)Cire\c{s}an, Meier, Masci, Gambardella, \& Schmidhuber}]{10.5555/2283516.2283603}
Cire\c{s}an, D.~C., Meier, U., Masci, J., Gambardella, L.~M., \& Schmidhuber, J. 2011, in Proceedings of the Twenty-Second International Joint Conference on Artificial Intelligence - Volume Volume Two, IJCAI'11 (AAAI Press), 1237–1242

\bibitem[{{Connerney} {et~al.}(2017){Connerney}, {Benn}, {Bjarno}, {Denver}, {Espley}, {Jorgensen}, {Jorgensen}, {Lawton}, {Malinnikova}, {Merayo}, {Murphy}, {Odom}, {Oliversen}, {Schnurr}, {Sheppard}, \& {Smith}}]{2017SSRv..213...39C}
{Connerney}, J.~E.~P., {Benn}, M., {Bjarno}, J.~B., {et~al.} 2017, \ssr, 213, 39, \dodoi{10.1007/s11214-017-0334-z}

\bibitem[{Cortes \& Vapnik(1995)}]{cortes1995support}
Cortes, C., \& Vapnik, V. 1995, Machine learning, 20, 273

\bibitem[{{Crooker} {et~al.}(1997){Crooker}, {Joselyn}, \& {Feynman}}]{1997GMS....99.....C}
{Crooker}, N., {Joselyn}, J.~A., \& {Feynman}, J. 1997, Geophysical Monograph Series, 99, \dodoi{10.1029/GM099}

\bibitem[{{Daglis}(2001)}]{2001SSRv...98..343D}
{Daglis}, I.~A. 2001, \ssr, 98, 343, \dodoi{10.1023/A:1013873329054}

\bibitem[{{Davies} {et~al.}(2021){Davies}, {Forsyth}, {Winslow}, {M{\"o}stl}, \& {Lugaz}}]{2021ApJ...923..136D}
{Davies}, E.~E., {Forsyth}, R.~J., {Winslow}, R.~M., {M{\"o}stl}, C., \& {Lugaz}, N. 2021, \apj, 923, 136, \dodoi{10.3847/1538-4357/ac2ccb}

\bibitem[{Deserno(2004)}]{deserno2004generate}
Deserno, M. 2004, If Polymerforshung (Ed.), 99, 1

\bibitem[{{dos Santos} {et~al.}(2020){dos Santos}, {Narock}, {Nieves-Chinchilla}, {Nu{\~n}ez}, \& {Kirk}}]{2020SoPh..295..131D}
{dos Santos}, L. F.~G., {Narock}, A., {Nieves-Chinchilla}, T., {Nu{\~n}ez}, M., \& {Kirk}, M. 2020, \solphys, 295, 131, \dodoi{10.1007/s11207-020-01697-x}

\bibitem[{{Dungey}(1961)}]{1961PhRvL...6...47D}
{Dungey}, J.~W. 1961, \prl, 6, 47, \dodoi{10.1103/PhysRevLett.6.47}

\bibitem[{{Fairfield} \& {Cahill}(1966)}]{1966JGR....71..155F}
{Fairfield}, D.~H., \& {Cahill}, L.~J., J. 1966, \jgr, 71, 155, \dodoi{10.1029/JZ071i001p00155}

\bibitem[{{Farooki} {et~al.}(2024){Farooki}, {Abduallah}, {Noh}, {Kim}, {Bizos}, {Shin}, {Wang}, \& {Wang}}]{2024ApJ...961...81F}
{Farooki}, H., {Abduallah}, Y., {Noh}, S.~J., {et~al.} 2024, \apj, 961, 81, \dodoi{10.3847/1538-4357/ad0c52}

\bibitem[{{Fox} {et~al.}(2016){Fox}, {Velli}, {Bale}, {Decker}, {Driesman}, {Howard}, {Kasper}, {Kinnison}, {Kusterer}, {Lario}, {Lockwood}, {McComas}, {Raouafi}, \& {Szabo}}]{2016SSRv..204....7F}
{Fox}, N.~J., {Velli}, M.~C., {Bale}, S.~D., {et~al.} 2016, \ssr, 204, 7, \dodoi{10.1007/s11214-015-0211-6}

\bibitem[{G{\'e}ron(2017)}]{geron2017handson}
G{\'e}ron, A. 2017, Hands-on machine learning with Scikit-Learn and TensorFlow : concepts, tools, and techniques to build intelligent systems (Sebastopol, CA: O'Reilly Media)

\bibitem[{{Gonzalez} \& {Echer}(2005)}]{2005GeoRL..3218103G}
{Gonzalez}, W.~D., \& {Echer}, E. 2005, \grl, 32, L18103, \dodoi{10.1029/2005GL023486}

\bibitem[{{Gonzalez} {et~al.}(1994){Gonzalez}, {Joselyn}, {Kamide}, {Kroehl}, {Rostoker}, {Tsurutani}, \& {Vasyliunas}}]{1994JGR....99.5771G}
{Gonzalez}, W.~D., {Joselyn}, J.~A., {Kamide}, Y., {et~al.} 1994, \jgr, 99, 5771, \dodoi{10.1029/93JA02867}

\bibitem[{{Good} {et~al.}(2020){Good}, {Kilpua}, {Ala-Lahti}, {Osmane}, {Bale}, \& {Zhao}}]{2020ApJ...900L..32G}
{Good}, S.~W., {Kilpua}, E.~K.~J., {Ala-Lahti}, M., {et~al.} 2020, \apjl, 900, L32, \dodoi{10.3847/2041-8213/abb021}

\bibitem[{{Good} {et~al.}(2023){Good}, {Rantala}, {Jylh{\"a}}, {Chen}, {M{\"o}stl}, \& {Kilpua}}]{2023ApJ...956L..30G}
{Good}, S.~W., {Rantala}, O.~K., {Jylh{\"a}}, A. S.~M., {et~al.} 2023, \apjl, 956, L30, \dodoi{10.3847/2041-8213/acfd1c}

\bibitem[{Goodfellow {et~al.}(2016)Goodfellow, Bengio, \& Courville}]{Goodfellow-et-al-2016}
Goodfellow, I., Bengio, Y., \& Courville, A. 2016, Deep Learning (MIT Press)

\bibitem[{{Gopalswamy}(2008)}]{2008JASTP..70.2078G}
{Gopalswamy}, N. 2008, Journal of Atmospheric and Solar-Terrestrial Physics, 70, 2078, \dodoi{10.1016/j.jastp.2008.06.010}

\bibitem[{{Gopalswamy} {et~al.}(2017){Gopalswamy}, {Yashiro}, {Akiyama}, \& {Xie}}]{2017SoPh..292...65G}
{Gopalswamy}, N., {Yashiro}, S., {Akiyama}, S., \& {Xie}, H. 2017, \solphys, 292, 65, \dodoi{10.1007/s11207-017-1080-9}

\bibitem[{{Gruet} {et~al.}(2018){Gruet}, {Chandorkar}, {Sicard}, \& {Camporeale}}]{2018SpWea..16.1882G}
{Gruet}, M.~A., {Chandorkar}, M., {Sicard}, A., \& {Camporeale}, E. 2018, Space Weather, 16, 1882, \dodoi{10.1029/2018SW001898}

\bibitem[{{Hahn} \& {Abel}(2011)}]{2011MNRAS.415.2101H}
{Hahn}, O., \& {Abel}, T. 2011, \mnras, 415, 2101, \dodoi{10.1111/j.1365-2966.2011.18820.x}

\bibitem[{{Hu} {et~al.}(2023){Hu}, {Camporeale}, \& {Swiger}}]{2023SpWea..2103286H}
{Hu}, A., {Camporeale}, E., \& {Swiger}, B. 2023, Space Weather, 21, e2022SW003286, \dodoi{10.1029/2022SW003286}

\bibitem[{{Hu} {et~al.}(2022){Hu}, {Shneider}, {Tiwari}, \& {Camporeale}}]{2022SpWea..2003064H}
{Hu}, A., {Shneider}, C., {Tiwari}, A., \& {Camporeale}, E. 2022, Space Weather, 20, e2022SW003064, \dodoi{10.1029/2022SW003064}

\bibitem[{{Hu} {et~al.}(2018){Hu}, {Zheng}, {Chen}, {le Roux}, \& {Zhao}}]{2018ApJS..239...12H}
{Hu}, Q., {Zheng}, J., {Chen}, Y., {le Roux}, J., \& {Zhao}, L. 2018, \apjs, 239, 12, \dodoi{10.3847/1538-4365/aae57d}

\bibitem[{{Jian} {et~al.}(2006){Jian}, {Russell}, {Luhmann}, \& {Skoug}}]{2006SoPh..239..393J}
{Jian}, L., {Russell}, C.~T., {Luhmann}, J.~G., \& {Skoug}, R.~M. 2006, \solphys, 239, 393, \dodoi{10.1007/s11207-006-0133-2}

\bibitem[{{Jian} {et~al.}(2018){Jian}, {Russell}, {Luhmann}, \& {Galvin}}]{2018ApJ...855..114J}
{Jian}, L.~K., {Russell}, C.~T., {Luhmann}, J.~G., \& {Galvin}, A.~B. 2018, \apj, 855, 114, \dodoi{10.3847/1538-4357/aab189}

\bibitem[{{Kahler} \& {Webb}(2007)}]{2007JGRA..112.9103K}
{Kahler}, S.~W., \& {Webb}, D.~F. 2007, Journal of Geophysical Research (Space Physics), 112, A09103, \dodoi{10.1029/2007JA012358}

\bibitem[{{Kaiser} {et~al.}(2008){Kaiser}, {Kucera}, {Davila}, {St. Cyr}, {Guhathakurta}, \& {Christian}}]{2008SSRv..136....5K}
{Kaiser}, M.~L., {Kucera}, T.~A., {Davila}, J.~M., {et~al.} 2008, \ssr, 136, 5, \dodoi{10.1007/s11214-007-9277-0}

\bibitem[{{Kane}(2005)}]{2005JGRA..110.2213K}
{Kane}, R.~P. 2005, Journal of Geophysical Research (Space Physics), 110, A02213, \dodoi{10.1029/2004JA010799}

\bibitem[{{Kilpua} {et~al.}(2009){Kilpua}, {Luhmann}, {Gosling}, {Li}, {Elliott}, {Russell}, {Jian}, {Galvin}, {Larson}, {Schroeder}, {Simunac}, \& {Petrie}}]{2009SoPh..256..327K}
{Kilpua}, E.~K.~J., {Luhmann}, J.~G., {Gosling}, J., {et~al.} 2009, \solphys, 256, 327, \dodoi{10.1007/s11207-009-9366-1}

\bibitem[{Kingma \& Ba(2014)}]{Kingma2014AdamAM}
Kingma, D.~P., \& Ba, J. 2014, CoRR, abs/1412.6980.
\newblock \url{https://api.semanticscholar.org/CorpusID:6628106}

\bibitem[{{Klein} \& {Burlaga}(1982)}]{1982JGR....87..613K}
{Klein}, L.~W., \& {Burlaga}, L.~F. 1982, \jgr, 87, 613, \dodoi{10.1029/JA087iA02p00613}

\bibitem[{{LeCun} {et~al.}(2015){LeCun}, {Bengio}, \& {Hinton}}]{2015Natur.521..436L}
{LeCun}, Y., {Bengio}, Y., \& {Hinton}, G. 2015, \nat, 521, 436, \dodoi{10.1038/nature14539}

\bibitem[{{Lepping} {et~al.}(2006){Lepping}, {Berdichevsky}, {Wu}, {Szabo}, {Narock}, {Mariani}, {Lazarus}, \& {Quivers}}]{2006AnGeo..24..215L}
{Lepping}, R.~P., {Berdichevsky}, D.~B., {Wu}, C.~C., {et~al.} 2006, Annales Geophysicae, 24, 215, \dodoi{10.5194/angeo-24-215-2006}

\bibitem[{{Lepping} {et~al.}(1995){Lepping}, {Ac{\~{u}}na}, {Burlaga}, {Farrell}, {Slavin}, {Schatten}, {Mariani}, {Ness}, {Neubauer}, {Whang}, {Byrnes}, {Kennon}, {Panetta}, {Scheifele}, \& {Worley}}]{1995SSRv...71..207L}
{Lepping}, R.~P., {Ac{\~{u}}na}, M.~H., {Burlaga}, L.~F., {et~al.} 1995, \ssr, 71, 207, \dodoi{10.1007/BF00751330}

\bibitem[{{Liu} {et~al.}(2018){Liu}, {Ye}, {Shen}, {Wang}, \& {Erd{\'e}lyi}}]{2018ApJ...855..109L}
{Liu}, J., {Ye}, Y., {Shen}, C., {Wang}, Y., \& {Erd{\'e}lyi}, R. 2018, \apj, 855, 109, \dodoi{10.3847/1538-4357/aaae69}

\bibitem[{{Lu} {et~al.}(2016){Lu}, {Peng}, {Wang}, {Gu}, \& {Zhao}}]{2016P&SS..120...48L}
{Lu}, J.~Y., {Peng}, Y.~X., {Wang}, M., {Gu}, S.~J., \& {Zhao}, M.~X. 2016, \planss, 120, 48, \dodoi{10.1016/j.pss.2015.11.004}

\bibitem[{{Lugaz} {et~al.}(2009){Lugaz}, {Vourlidas}, \& {Roussev}}]{2009AnGeo..27.3479L}
{Lugaz}, N., {Vourlidas}, A., \& {Roussev}, I.~I. 2009, Annales Geophysicae, 27, 3479, \dodoi{10.5194/angeo-27-3479-2009}

\bibitem[{{Luhmann} {et~al.}(2020){Luhmann}, {Gopalswamy}, {Jian}, \& {Lugaz}}]{2020SoPh..295...61L}
{Luhmann}, J.~G., {Gopalswamy}, N., {Jian}, L.~K., \& {Lugaz}, N. 2020, \solphys, 295, 61, \dodoi{10.1007/s11207-020-01624-0}

\bibitem[{{Luhmann} {et~al.}(2008){Luhmann}, {Curtis}, {Schroeder}, {McCauley}, {Lin}, {Larson}, {Bale}, {Sauvaud}, {Aoustin}, {Mewaldt}, {Cummings}, {Stone}, {Davis}, {Cook}, {Kecman}, {Wiedenbeck}, {von Rosenvinge}, {Acuna}, {Reichenthal}, {Shuman}, {Wortman}, {Reames}, {Mueller-Mellin}, {Kunow}, {Mason}, {Walpole}, {Korth}, {Sanderson}, {Russell}, \& {Gosling}}]{2008SSRv..136..117L}
{Luhmann}, J.~G., {Curtis}, D.~W., {Schroeder}, P., {et~al.} 2008, \ssr, 136, 117, \dodoi{10.1007/s11214-007-9170-x}

\bibitem[{{Manchester} {et~al.}(2017){Manchester}, {Kilpua}, {Liu}, {Lugaz}, {Riley}, {T{\"o}r{\"o}k}, \& {Vr{\v{s}}nak}}]{2017SSRv..212.1159M}
{Manchester}, W., {Kilpua}, E. K.~J., {Liu}, Y.~D., {et~al.} 2017, \ssr, 212, 1159, \dodoi{10.1007/s11214-017-0394-0}

\bibitem[{{Moldwin} {et~al.}(2000){Moldwin}, {Ford}, {Lepping}, {Slavin}, \& {Szabo}}]{2000GeoRL..27...57M}
{Moldwin}, M.~B., {Ford}, S., {Lepping}, R., {Slavin}, J., \& {Szabo}, A. 2000, \grl, 27, 57, \dodoi{10.1029/1999GL010724}

\bibitem[{{Moldwin} {et~al.}(1995){Moldwin}, {Phillips}, {Gosling}, {Scime}, {McComas}, {Bame}, {Balogh}, \& {Forsyth}}]{1995JGR...10019903M}
{Moldwin}, M.~B., {Phillips}, J.~L., {Gosling}, J.~T., {et~al.} 1995, \jgr, 100, 19903, \dodoi{10.1029/95JA01123}

\bibitem[{Mucherino {et~al.}(2009)Mucherino, Papajorgji, \& Pardalos}]{Mucherino2009}
Mucherino, A., Papajorgji, P.~J., \& Pardalos, P.~M. 2009, k-Nearest Neighbor Classification (New York, NY: Springer New York), 83--106, \dodoi{10.1007/978-0-387-88615-2_4}

\bibitem[{{M{\"u}ller} {et~al.}(2020){M{\"u}ller}, {St. Cyr}, {Zouganelis}, {Gilbert}, {Marsden}, {Nieves-Chinchilla}, {Antonucci}, {Auch{\`e}re}, {Berghmans}, {Horbury}, {Howard}, {Krucker}, {Maksimovic}, {Owen}, {Rochus}, {Rodriguez-Pacheco}, {Romoli}, {Solanki}, {Bruno}, {Carlsson}, {Fludra}, {Harra}, {Hassler}, {Livi}, {Louarn}, {Peter}, {Sch{\"u}hle}, {Teriaca}, {del Toro Iniesta}, {Wimmer-Schweingruber}, {Marsch}, {Velli}, {De Groof}, {Walsh}, \& {Williams}}]{2020A&A...642A...1M}
{M{\"u}ller}, D., {St. Cyr}, O.~C., {Zouganelis}, I., {et~al.} 2020, \aap, 642, A1, \dodoi{10.1051/0004-6361/202038467}

\bibitem[{Nair {et~al.}(2023)Nair, Redmon, Young, Chulliat, Trotta, Chung, Lipstein, \& Slavitt}]{nair2023magnet}
Nair, M., Redmon, R., Young, L.-Y., {et~al.} 2023, Space Weather, 21, e2023SW003514

\bibitem[{Nair \& Hinton(2010)}]{Nair2010RectifiedLU}
Nair, V., \& Hinton, G.~E. 2010, in International Conference on Machine Learning.
\newblock \url{https://api.semanticscholar.org/CorpusID:15539264}

\bibitem[{{Nandy} {et~al.}(2023){Nandy}, {Baruah}, {Bhowmik}, {Dash}, {Gupta}, {Hazra}, {Lekshmi}, {Pal}, {Pal}, {Roy}, {Saha}, \& {Sinha}}]{2023JASTP.24806081N}
{Nandy}, D., {Baruah}, Y., {Bhowmik}, P., {et~al.} 2023, Journal of Atmospheric and Solar-Terrestrial Physics, 248, 106081, \dodoi{10.1016/j.jastp.2023.106081}

\bibitem[{{Narock} {et~al.}(2022){Narock}, {Narock}, {Dos Santos}, \& {Nieves-Chinchilla}}]{2022FrASS...938442N}
{Narock}, T., {Narock}, A., {Dos Santos}, L. F.~G., \& {Nieves-Chinchilla}, T. 2022, Frontiers in Astronomy and Space Sciences, 9, 838442, \dodoi{10.3389/fspas.2022.838442}

\bibitem[{{Nguyen} {et~al.}(2019){Nguyen}, {Aunai}, {Fontaine}, {Le Pennec}, {Van den Bossche}, {Jeandet}, {Bakkali}, {Vignoli}, \& {Regaldo-Saint Blancard}}]{2019ApJ...874..145N}
{Nguyen}, G., {Aunai}, N., {Fontaine}, D., {et~al.} 2019, \apj, 874, 145, \dodoi{10.3847/1538-4357/ab0d24}

\bibitem[{{Nieves-Chinchilla} {et~al.}(2023){Nieves-Chinchilla}, {Hidalgo}, \& {Cremades}}]{2023ApJ...947...79N}
{Nieves-Chinchilla}, T., {Hidalgo}, M.~A., \& {Cremades}, H. 2023, \apj, 947, 79, \dodoi{10.3847/1538-4357/acb3c1}

\bibitem[{{Nieves-Chinchilla} {et~al.}(2019){Nieves-Chinchilla}, {Jian}, {Balmaceda}, {Vourlidas}, {dos Santos}, \& {Szabo}}]{2019SoPh..294...89N}
{Nieves-Chinchilla}, T., {Jian}, L.~K., {Balmaceda}, L., {et~al.} 2019, \solphys, 294, 89, \dodoi{10.1007/s11207-019-1477-8}

\bibitem[{{Nieves-Chinchilla} {et~al.}(2016){Nieves-Chinchilla}, {Linton}, {Hidalgo}, {Vourlidas}, {Savani}, {Szabo}, {Farrugia}, \& {Yu}}]{2016ApJ...823...27N}
{Nieves-Chinchilla}, T., {Linton}, M.~G., {Hidalgo}, M.~A., {et~al.} 2016, \apj, 823, 27, \dodoi{10.3847/0004-637X/823/1/27}

\bibitem[{{Nieves-Chinchilla} {et~al.}(2018){Nieves-Chinchilla}, {Vourlidas}, {Raymond}, {Linton}, {Al-haddad}, {Savani}, {Szabo}, \& {Hidalgo}}]{2018SoPh..293...25N}
{Nieves-Chinchilla}, T., {Vourlidas}, A., {Raymond}, J.~C., {et~al.} 2018, \solphys, 293, 25, \dodoi{10.1007/s11207-018-1247-z}

\bibitem[{{Nitta} {et~al.}(2021){Nitta}, {Mulligan}, {Kilpua}, {Lynch}, {Mierla}, {O'Kane}, {Pagano}, {Palmerio}, {Pomoell}, {Richardson}, {Rodriguez}, {Rouillard}, {Sinha}, {Srivastava}, {Talpeanu}, {Yardley}, \& {Zhukov}}]{Nitta2021}
{Nitta}, N.~V., {Mulligan}, T., {Kilpua}, E. K.~J., {et~al.} 2021, \ssr, 217, 82, \dodoi{10.1007/s11214-021-00857-0}

\bibitem[{{Pal}(2022)}]{2022AdSpR..70.1601P}
{Pal}, S. 2022, Advances in Space Research, 70, 1601, \dodoi{10.1016/j.asr.2021.11.013}

\bibitem[{{Pal} {et~al.}(2023){Pal}, {Balmaceda}, {Weiss}, {Nieves-Chinchilla}, {Carcaboso}, {Kilpua}, \& {M{\"o}stl}}]{2023FrASS..1095805P}
{Pal}, S., {Balmaceda}, L., {Weiss}, A.~J., {et~al.} 2023, Frontiers in Astronomy and Space Sciences, 10, 1195805, \dodoi{10.3389/fspas.2023.1195805}

\bibitem[{{Pal} {et~al.}(2017){Pal}, {Gopalswamy}, {Nandy}, {Akiyama}, {Yashiro}, {Makela}, \& {Xie}}]{2017ApJ...851..123P}
{Pal}, S., {Gopalswamy}, N., {Nandy}, D., {et~al.} 2017, \apj, 851, 123, \dodoi{10.3847/1538-4357/aa9983}

\bibitem[{{Pal} {et~al.}(2022){Pal}, {Nandy}, \& {Kilpua}}]{2022A&A...665A.110P}
{Pal}, S., {Nandy}, D., \& {Kilpua}, E. K.~J. 2022, \aap, 665, A110, \dodoi{10.1051/0004-6361/202243513}

\bibitem[{{Pal} {et~al.}(2018){Pal}, {Nandy}, {Srivastava}, {Gopalswamy}, \& {Panda}}]{2018ApJ...865....4P}
{Pal}, S., {Nandy}, D., {Srivastava}, N., {Gopalswamy}, N., \& {Panda}, S. 2018, \apj, 865, 4, \dodoi{10.3847/1538-4357/aada10}

\bibitem[{Pedregosa {et~al.}(2011)Pedregosa, Varoquaux, Gramfort, Michel, Thirion, Grisel, Blondel, Prettenhofer, Weiss, Dubourg, {et~al.}}]{pedregosa2011scikit}
Pedregosa, F., Varoquaux, G., Gramfort, A., {et~al.} 2011, the Journal of machine Learning research, 12, 2825

\bibitem[{{Pricopi} {et~al.}(2022){Pricopi}, {Paraschiv}, {Besliu-Ionescu}, \& {Marginean}}]{2022ApJ...934..176P}
{Pricopi}, A.-C., {Paraschiv}, A.~R., {Besliu-Ionescu}, D., \& {Marginean}, A.-N. 2022, \apj, 934, 176, \dodoi{10.3847/1538-4357/ac7962}

\bibitem[{{Qahwaji} \& {Colak}(2007)}]{2007SoPh..241..195Q}
{Qahwaji}, R., \& {Colak}, T. 2007, \solphys, 241, 195, \dodoi{10.1007/s11207-006-0272-5}

\bibitem[{{Qiu} {et~al.}(2007){Qiu}, {Hu}, {Howard}, \& {Yurchyshyn}}]{2007ApJ...659..758Q}
{Qiu}, J., {Hu}, Q., {Howard}, T.~A., \& {Yurchyshyn}, V.~B. 2007, \apj, 659, 758, \dodoi{10.1086/512060}

\bibitem[{{Richardson} \& {Cane}(1995)}]{1995JGR...10023397R}
{Richardson}, I.~G., \& {Cane}, H.~V. 1995, \jgr, 100, 23397, \dodoi{10.1029/95JA02684}

\bibitem[{{Richardson} \& {Cane}(2004)}]{2004GeoRL..3118804R}
---. 2004, \grl, 31, L18804, \dodoi{10.1029/2004GL020958}

\bibitem[{{Richardson} \& {Cane}(2010)}]{2010SoPh..264..189R}
---. 2010, \solphys, 264, 189, \dodoi{10.1007/s11207-010-9568-6}

\bibitem[{{Rodr{\'\i}guez} {et~al.}(2022){Rodr{\'\i}guez}, {Rodr{\'\i}guez-Rodr{\'\i}guez}, \& {Lok Woo}}]{2022PASP..134l4201R}
{Rodr{\'\i}guez}, J.-V., {Rodr{\'\i}guez-Rodr{\'\i}guez}, I., \& {Lok Woo}, W. 2022, \pasp, 134, 124201, \dodoi{10.1088/1538-3873/aca4a3}

\bibitem[{Ronneberger {et~al.}(2015)Ronneberger, Fischer, \& Brox}]{10.1007/978-3-319-24574-4_28}
Ronneberger, O., Fischer, P., \& Brox, T. 2015, in Medical Image Computing and Computer-Assisted Intervention -- MICCAI 2015, ed. N.~Navab, J.~Hornegger, W.~M. Wells, \& A.~F. Frangi (Cham: Springer International Publishing), 234--241

\bibitem[{{R{\"u}disser} {et~al.}(2022){R{\"u}disser}, {Windisch}, {Amerstorfer}, {M{\"o}stl}, {Amerstorfer}, {Bailey}, \& {Reiss}}]{2022SpWea..2003149R}
{R{\"u}disser}, H.~T., {Windisch}, A., {Amerstorfer}, U.~V., {et~al.} 2022, Space Weather, 20, e2022SW003149, \dodoi{10.1029/2022SW003149}

\bibitem[{{Russell} \& {Shinde}(2003)}]{2003SoPh..216..285R}
{Russell}, C.~T., \& {Shinde}, A.~A. 2003, \solphys, 216, 285, \dodoi{10.1023/A:1026108101883}

\bibitem[{{Salman} {et~al.}(2024){Salman}, {Nieves-Chinchilla}, {Jian}, {Lugaz}, {Carcaboso}, {Davies}, \& {Collado-Vega}}]{2024ApJ...966..118S}
{Salman}, T.~M., {Nieves-Chinchilla}, T., {Jian}, L.~K., {et~al.} 2024, \apj, 966, 118, \dodoi{10.3847/1538-4357/ad320c}

\bibitem[{{Schekochihin}(2022)}]{2022JPlPh..88e1501S}
{Schekochihin}, A.~A. 2022, Journal of Plasma Physics, 88, 155880501, \dodoi{10.1017/S0022377822000721}

\bibitem[{{Sheeley} {et~al.}(2008){Sheeley}, {Herbst}, {Palatchi}, {Wang}, {Howard}, {Moses}, {Vourlidas}, {Newmark}, {Socker}, {Plunkett}, {Korendyke}, {Burlaga}, {Davila}, {Thompson}, {St. Cyr}, {Harrison}, {Davis}, {Eyles}, {Halain}, {Wang}, {Rich}, {Battams}, {Esfandiari}, \& {Stenborg}}]{2008ApJ...675..853S}
{Sheeley}, N.~R., J., {Herbst}, A.~D., {Palatchi}, C.~A., {et~al.} 2008, \apj, 675, 853, \dodoi{10.1086/526422}

\bibitem[{{Sheeley} {et~al.}(1997){Sheeley}, {Wang}, {Hawley}, {Brueckner}, {Dere}, {Howard}, {Koomen}, {Korendyke}, {Michels}, {Paswaters}, {Socker}, {St. Cyr}, {Wang}, {Lamy}, {Llebaria}, {Schwenn}, {Simnett}, {Plunkett}, \& {Biesecker}}]{1997ApJ...484..472S}
{Sheeley}, N.~R., {Wang}, Y.~M., {Hawley}, S.~H., {et~al.} 1997, \apj, 484, 472, \dodoi{10.1086/304338}

\bibitem[{{Shinde} \& {Russell}(2003)}]{2003AGUFMSH21B0133S}
{Shinde}, A.~A., \& {Russell}, C.~T. 2003, in AGU Fall Meeting Abstracts, Vol. 2003, SH21B--0133

\bibitem[{{Sinha} {et~al.}(2022){Sinha}, {Gupta}, {Singh}, {Lekshmi}, {Nandy}, {Mitra}, {Chatterjee}, {Bhattacharya}, {Chatterjee}, {Srivastava}, {Brandenburg}, \& {Pal}}]{2022ApJ...935...45S}
{Sinha}, S., {Gupta}, O., {Singh}, V., {et~al.} 2022, \apj, 935, 45, \dodoi{10.3847/1538-4357/ac7955}

\bibitem[{{Sun} {et~al.}(2020){Sun}, {Slavin}, {Smith}, {Dewey}, {Poh}, {Jia}, {Raines}, {Livi}, {Saito}, {Gershman}, {DiBraccio}, {Imber}, {Guo}, {Fu}, {Zong}, \& {Zhao}}]{2020EPSC...14...62S}
{Sun}, W.-J., {Slavin}, J., {Smith}, A., {et~al.} 2020, in European Planetary Science Congress, EPSC2020--62, \dodoi{10.5194/epsc2020-62}

\bibitem[{Taunk {et~al.}(2019)Taunk, De, Verma, \& Swetapadma}]{9065747}
Taunk, K., De, S., Verma, S., \& Swetapadma, A. 2019, in 2019 International Conference on Intelligent Computing and Control Systems (ICCS), 1255--1260, \dodoi{10.1109/ICCS45141.2019.9065747}

\bibitem[{{Vourlidas} \& {Webb}(2018)}]{Vourlidas2018}
{Vourlidas}, A., \& {Webb}, D.~F. 2018, \apj, 861, 103, \dodoi{10.3847/1538-4357/aaca3e}

\bibitem[{{Wang} {et~al.}(2003){Wang}, {Shen}, {Wang}, \& {Ye}}]{2003GeoRL..30.2039W}
{Wang}, Y., {Shen}, C.~L., {Wang}, S., \& {Ye}, P.~Z. 2003, \grl, 30, 2039, \dodoi{10.1029/2003GL017901}

\bibitem[{{Wang} {et~al.}(2000){Wang}, {Sheeley}, {Socker}, {Howard}, \& {Rich}}]{2000JGR...10525133W}
{Wang}, Y.~M., {Sheeley}, N.~R., {Socker}, D.~G., {Howard}, R.~A., \& {Rich}, N.~B. 2000, \jgr, 105, 25133, \dodoi{10.1029/2000JA000149}

\bibitem[{{Webb} \& {Howard}(2012)}]{Webb_Howard2012}
{Webb}, D.~F., \& {Howard}, T.~A. 2012, LRSP, 9, 3, \dodoi{10.12942/lrsp-2012-3}

\bibitem[{{Weiss} {et~al.}(2022){Weiss}, {Nieves-Chinchilla}, {M{\"o}stl}, {Reiss}, {Amerstorfer}, \& {Bailey}}]{2022JGRA..12730898W}
{Weiss}, A.~J., {Nieves-Chinchilla}, T., {M{\"o}stl}, C., {et~al.} 2022, Journal of Geophysical Research (Space Physics), 127, e2022JA030898, \dodoi{10.1029/2022JA030898}

\bibitem[{{Weiss} {et~al.}(2021){Weiss}, {M{\"o}stl}, {Davies}, {Amerstorfer}, {Bauer}, {Hinterreiter}, {Reiss}, {Bailey}, {Horbury}, {O'Brien}, {Evans}, {Angelini}, {Heyner}, {Richter}, {Auster}, {Magnes}, {Fischer}, \& {Baumjohann}}]{2021A&A...656A..13W}
{Weiss}, A.~J., {M{\"o}stl}, C., {Davies}, E.~E., {et~al.} 2021, \aap, 656, A13, \dodoi{10.1051/0004-6361/202140919}

\bibitem[{{Welch}(1967)}]{1967ITAE...15...70W}
{Welch}, P.~D. 1967, IEEE Trans. Audio \& Electroacoust, 15, 70

\bibitem[{{Wu} \& {Lepping}(2002)}]{2002JGRA..107.1314W}
{Wu}, C.-C., \& {Lepping}, R.~P. 2002, Journal of Geophysical Research (Space Physics), 107, 1314, \dodoi{10.1029/2001JA000161}

\bibitem[{{Xu} {et~al.}(2020){Xu}, {Huang}, {Yuan}, {Deng}, \& {Jiang}}]{2020ApJS..248...14X}
{Xu}, S.~B., {Huang}, S.~Y., {Yuan}, Z.~G., {Deng}, X.~H., \& {Jiang}, K. 2020, \apjs, 248, 14, \dodoi{10.3847/1538-4365/ab880e}

\bibitem[{{Yu} {et~al.}(2016){Yu}, {Farrugia}, {Galvin}, {Lugaz}, {Luhmann}, {Simunac}, \& {Kilpua}}]{2016JGRA..121.5005Y}
{Yu}, W., {Farrugia}, C.~J., {Galvin}, A.~B., {et~al.} 2016, Journal of Geophysical Research (Space Physics), 121, 5005, \dodoi{10.1002/2016JA022642}

\bibitem[{{Yu} {et~al.}(2014){Yu}, {Farrugia}, {Lugaz}, {Galvin}, {Kilpua}, {Kucharek}, {M{\"o}stl}, {Leitner}, {Torbert}, {Simunac}, {Luhmann}, {Szabo}, {Wilson}, {Ogilvie}, \& {Sauvaud}}]{2014JGRA..119..689Y}
{Yu}, W., {Farrugia}, C.~J., {Lugaz}, N., {et~al.} 2014, Journal of Geophysical Research (Space Physics), 119, 689, \dodoi{10.1002/2013JA019115}

\end{thebibliography}
\end{document}